\documentclass[12pt]{article}
\usepackage{amsfonts}

\font\mybb=msbm10 at 12pt
\def\bb#1{\hbox{\mybb#1}}
\renewcommand{\theequation}{\arabic{section}.\arabic{equation}}

\begin{document}

\thispagestyle{empty}

\begin{flushright}
December 11, 2009 
\end{flushright}

\vskip 1.5cm

\begin{center}
\baselineskip=16pt {\LARGE\bf Superembedding approach to Dp-branes, M-branes and multiple D(0)--brane systems}

\vskip 1cm

 {\large\bf Igor A. Bandos\footnote{Also at Institute for Theoretical Physcs,
NSC Kharkov Institute of Physics \& Technology,
UA 61108,  Kharkov, Ukraine. E-mail: igor\_bandos@ehu.es, bandos@ific.uv.es }}
\vskip 1.0cm {\it\small $^\dagger$ IKERBASQUE, the Basque Foundation for Science, and  \\ Department of Theoretical Physics,
University of the Basque Country,  \\ P.O. Box 644, 48080 Bilbao,
Spain
}
  \vspace{12pt}

\end{center}

\vskip 0.5cm

\par
\begin{quote}

We review the superembedding approach to M-branes and D$p$-branes, in its form based on the universal (D and p-independent) {\it superembedding equation},  and its recent application in searching for supersymmetric and Lorentz covariant description of multiple D$p$--brane systems. In particular, we present the structure of the multiple D0-brane equation as follows from our superembedding description  and show that it describes the dielectric effect firstly noticed by Emparan and then by Myers. We also discuss briefly  the relation with the boundary fermion approach by Howe, Lindstr\"om and Wulff.

\end{quote}

\newpage

\setcounter{page}{2}
\section{Introduction}
\setcounter{equation}{0}

Supersymmetric extended objects, super--$p$--branes \cite{BST87,AETW,hs96,Dpac,hs2,Dpac2,B+T=Dpac,blnpst,schw5}, play a very important r\^ole in String/M--theory \cite{Duality,M-theory} and its ADS/CFT applications \cite{AdS/CFT,AdS/CFTrev}.  The ground states of a $D$--dimensional  super-$p$-branes (superstring for $p=1$, supermembrane for $p=2$) can be identified with the supersymmetric solutions of the corresponding supergravity theories \cite{SUGRA}.
The most interesting are the solutions of the maximal $D=11$ supergravity and type II $D=10$ supergravities appearing as low energy limit of type II superstring theories.
The $p$-brane dynamics can be described by supersymmetric actions  \cite{BST87,AETW,Dpac,Dpac2,B+T=Dpac,blnpst,schw5} or in the frame of superembedding approach \cite{bpstv,hs96,hs2,bst97,HS+Chu=PLB98,ABKZ,Dima99}.

In this contribution we give a review of superembedding approach to super--$p$--branes \cite{bpstv,hs96,hs2,bst97,HS+Chu=PLB98,ABKZ,Dima99,B00,IB08:Q7} in D=10 and D=11 superspaces and its recent application in search for the supersymmetric and Lorentz covariant (diffeomorphism invariant) description of the multiple brane systems \cite{IB09:D0}.  In the part devoted to superembedding description of a single brane  our emphasis will be on the superembedding description of Dirichlet super--$p$--branes (D$p$--branes) (in contrast with  the already existing review \cite{Dima99}). We begin by this case and then turn to the superembedding description of M2- and  M5-brane.

The part devoted to multiple branes  contains the results on multiple D$0$--brane system, which is to say multiple D-particles, which were briefly reported in \cite{IB09:D0}. We argue that, to describe the multiple D$p$-brane system, it is natural to try to put an additional  $SU(N)$ gauge superform on the worldvolume of a single D$p$-brane and impose a suitable set of superspace constraints on their  (super)field strength 2-form. If consistent, such a system provides, at least, an approximation of nearly coincident D$p$-brane with the very low energy of the relative motion, but with the  nonlinear  (`complete') Dirac-Born-Infeld  description of the dynamics of the center of mass and of the  $U(1)$ gauge field related to it. We show that such a consistent description, going beyond $U(N)$ Super-Yang-Mills (or Matrix model) approximation does exist at least for the case of multiple D$0$-brane system \cite{IB09:D0}.
We discuss the structure of the multiple D0-brane equation which follows from the superembedding approach and show that it possesses the dielectric effect firstly noticed by Emparan \cite{Emparan:1997rt} and then by Myers \cite{Myers:1999ps}.

Discussing the meaning of our results, we describe possible deformation of our  basic equations and the relation with the boundary fermion approach by Howe, Lindstr\"om and Wulff  \cite{Howe+Linstrom+Linus,Howe+Linstrom+Linus=2007}. This latter approach does provide supersymmetric and covariant description of Dirichlet branes, but on the 'classical' (or 'menus one quantization' level) in the sense that to arrive at the description of multiple brane system in terms of the variables corresponding to the standard single D$p$--brane action \cite{Dpac,Dpac2,B+T=Dpac} (usually considered as a classical or quasi-classical action) one has to perform a quantization of the boundary fermion sector.

\subsection{D-branes and multiple D-brane systems}

The first appearance of D-branes (Dirichlet $p$-branes) is dated by late 80th, when they were found as surfaces where the fundamental string can end \cite{Sagnotti,Horava,Polchinski89,Leigh:1989jq}. Although in the first quantized string model they appeared as flat hyperplanes, it was clear that these surfaces must be dynamical in string theory. Indeed, as far as the open string theory contains closed string sector and this contains gravity in its quantum state spectrum, nondynamical surfaces cannot exist in String theory as the spacetime itself is dynamical in it.

However, the  special importance of D-branes for String/M-theory \cite{Duality,M-theory} was widely appreciated in middle 90th, after it was discovered \cite{Polchinski95} that D$p$-branes carry  Ramond-Ramond (RR) charges {\it i.e.} that they interact with the antisymmetric tensor gauge fields $C_{p+1}$, $C_{p-1}$, $\ldots$ with  respect to which the fundamental strings is neutral. In particular, this made clear that D$p$--branes are described by supersymmetric $p$-brane solutions of extended ${\cal N}=2$ (type II) D=10 supergravity, which had been found for any even/odd value of $p$ in type IIA/IIB case and included a nonvanishing solution for $C_{p+1}$ RR gauge field equations.

It was quickly appreciated that the  low energy
dynamics of multiple Dp-brane system is described by the maximal
supersymmetric $d=p+1$ gauge theory with the gauge group $U(N)$ in
the case of N D-branes \cite{Witten:1995im}. The investigation of this limit was already
quite productive \cite{D0-1996}. In particular, it allowed to formulate the
conjecture of M(atrix) theory which states that the  Matrix model
\cite{Banks:1996vh}, which can be considered as a theory of multiple D0-brane
system, could provide a nonperturbative description of the M-theory.

The nonlinear supersymmetric action for a single
D$p$-brane was constructed in \cite{Townsend95} for $p=2$ and in
\cite{Dpac,Dpac2,B+T=Dpac} for general $p\;$ \footnote{The D-brane actions of  \cite{Townsend95} and
\cite{Dpac,Dpac2,B+T=Dpac} are complete up to terms containing the derivative of gauge field strength; in other words, they   include nonlinear effects but contain contributions of
lowest order in the derivatives of the field strength of the worldvolume gauge field only. Higher derivative corrections to this DBI+WZ actions are expected
 \cite{D-braneAnom}.}. It contains the nonlinear
Dirac--Born--Infeld (DBI) term \cite{Witten:1995im,Townsend95,Tseytlin:BI-DBI} and the Wess--Zumino (WZ) term  describing the coupling to RR gauge
fields $C_{p+1}$, $C_{p-1}$, $\ldots$ \cite{D-braneWZ}. Notice that this explains why, {\it e.g.} the odd $p$ D$p$-branes cannot exist in type IIA case, where the supergravity multiplet contains only odd form gauge potential, $C_{2n+1}$ which can be coupled to even $p$ super-$p$-branes (with odd dimension $d=p+1$ of the worldvolume $W^{p+1}$) through $\int_{W^{p+1}}\hat{C}_{p+1}$ (where hat implies pull--back of the differential form to $W^{p+1}$, see secs. 1.3 and 2.1 for the notation).

Even before the actions for generic D$p$-branes were constructed in
\cite{Dpac,Dpac2,B+T=Dpac}, the supersymmetric equations of motion were
derived in \cite{hs96} by developing superembedding approach
\cite{bpstv} for the case of Dp-branes. Notice that the same story happened with M5-brane: its equations of motion had been derived in
\cite{hs2} before the covariant and supersymmetric action was
constructed in \cite{blnpst} and, independently, in \cite{schw5}.

As far as the nonlinear action for multiple D-brane systems is
concerned, it was expected that this should be described by some
non-Abelian generalization of the DBI plus WZ  action.  Tseytlin
proposed to use the symmetric trace prescription to construct the
non-Abelian DBI  action for the case of purely bosonic spacetime
filling D-brane \cite{Tseytlin:DBInA,Tseytlin:BI-DBI}.

Although the search for a supersymmetric generalization of such
non-Abelian DBI action has  not been successful, in 1999 Myers used
it as a starting point and, applying a chain of dualities, derived
the so-called 'dielectric brane action' \cite{Myers:1999ps} which is
widely  accepted for the description of multiple D-brane system.
This action, however, does not possess neither supersymmetry nor
Lorentz symmetry. In spite of a number of attempts, its Lorentz
covariant and/or supersymmetric generalizations is not known in
general, although some progress was reached for the cases of low
dimensions $D$, low dimensional and low co-dimensional branes
\cite{Dima01,Drummond:2002kg}.

In \cite{Howe+Linstrom+Linus,Howe+Linstrom+Linus=2007} a very interesting, Lorentz covariant and supersymmetric description of D-branes is given in the frame of boundary fermion approach. It implies the extension of spacetime/superspace by new fermionic coordinates of the type introduced in \cite{Marcus:1986cm} as fields leaving at the end point of the open string. Upon quantization the boundary fermions of \cite{Marcus:1986cm} are replaced by Dirac matrices and reproduce the Chan-Paton factors in the open string amplitudes. In the approach of  \cite{Howe+Linstrom+Linus,Howe+Linstrom+Linus=2007} one also have to quantize the boundary fermion sector to arrive at the description of multiple D$p$--brane system similar to the standard description of single D$p$--brane in \cite{hs96,Dpac,Dpac2,B+T=Dpac}. In this sense the approach of \cite{Howe+Linstrom+Linus,Howe+Linstrom+Linus=2007} can be called {\it minus one quantization} of D$p$--brane. We will comment more on this approach in the concluding section of our review.

\medskip

As far as the superembedding approach shown its efficiency in
derivation of Dp-brane and M5-brane equations, it looks
natural to apply it in the search for equations of motion for the
multiple D$p$-brane system. In this review we describe the results which this procedure gives  for the simplest case of multiple D$0$-brane system \cite{IB09:D0}.\footnote{Notice that the boundary fermion approach \cite{Howe+Linstrom+Linus,Howe+Linstrom+Linus=2007} also uses a kind of superembedding formalism, but with embedding of a superspace with boundary fermion directions into spacetime \cite{Howe+Linstrom+Linus} or into the standard superspace  \cite{Howe+Linstrom+Linus=2007}.
Thus for a sufficiently large $N$ the number of fermionic directions of the worldvolume superspace exceeds $32$, which is the fermionic dimension of the target type II superspace $\Sigma^{(10|16+16)}$.
In this respect the boundary fermion approach is similar to  the superfield description of the NSR (Nevieu--Schwarz--Ramond) or spinning string, where the worldsheet superspace with two fermionic directions is embedded into spacetime (zero fermionic directions). In this review we are dealing with the standard superembedding approach, in which the worldvolume superspace has twice less fermionic directions than the target superspace ($16$ versus $32$ for 10 dimensional D-branes and 11 dimensional M-branes).}

\medskip

\subsection{Contents }

This review is organized as follows.
After establishing our basic notation  (a secs. 1.3 and 2.1), we begin (in Sec. 2) by describing the basic equations of the superembedding approach including the {\it superembedding equation} which essentially determines the dynamics of M-branes and D-branes (for a sufficiently large co-dimension $D-p> 4$).

In Sec. 3 we give  a very brief review of superembedding approach to single D$p$-branes for
arbitrary $p$, with particular emphasis on D$0$-brane case. In Sec. 4 we describe a more complicated cases of M2- and M5-brane where the construction of superembedding approach inevitably involves introduction of spinor moving frame variables (spinor harmonics) in additional to the moving frame variables.  In sec. 5 we first argue in favor of the  idea
to search for the description of multiple D$p$-brane systems by
trying to define a possible nonlinear generalization of the
non-Abelian SYM multiplet by some set of constraints on the
D$p$-brane worldvolume superspace ${\cal W}^{(p+1|16)}$ the embedding of which in the type II target superspace $\Sigma^{(10|32)}$ is determined by  superembedding equation.

Then, turning to the case of multiple D0-brane, we propose the d=1
${\cal N}=16$ SYM constraints which express its field strength in
terms of nanoplet of $su(N)$ valued superfields ${\bb X}^i$ obeying
a superembedding--like equation $D_\alpha {\bb
X}^i=(\sigma^{0i}\Psi)_\alpha$. The leading component of this
superfield, appearing in the expression for the dimension 1
(spinor-spinor) field strength of the $SU(N)$ gauge (super)fields,
$G_{\alpha\beta}=\sigma^i_{\alpha\beta}{\bb X}^i$, describes the
relative motion of N D$0$-brane constituents of the system. We show
that our constraints lead to interacting supersymmetric  equations of motion, which, in the
case of flat target superspace, can be also obtained by dimensional reduction of a non-Abelian D=10 Super Yang--Mills (SYM) theory to $d=1$ (the system which was used to define
the Matrix model).

However, the superembedding approach is also able
to produce multiple D0-brane equations in an arbitrary type IIA
superspace supergravity background (and, to our best knowledge, it is not clear how to
reproduce these equations just by SYM dimensional reduction). We
analyze the general algebraic structure of the bosonic equations of
motion for the multiple D0-brane in general type IIA supergravity
background, which follow from our superembedding approach, and show that
these describe the Emparan--Myers 'dielectric brane' effect \cite{Emparan:1997rt,Myers:1999ps} of polarization of multiple D$p$-brane system by external higher form fluxes,  {\it
i.e.} shows the  coupling of multiple D0-brane system to the higher
form gauge fields, which do not interact with a single D$0$-brane.

We conclude by discussion on our results, on possible generalizations of our approach and its relation with the boundary fermion approach by Howe, Lindstr\"om and Wulff \cite{Howe+Linstrom+Linus,Howe+Linstrom+Linus=2007}, and also on interesting directions
for future study.

\subsection{Basic notations}
\subsubsection{Target superspaces of D-branes and M-branes}
We denote the local coordinates of D=11 and type II D=10 superspace by
\begin{eqnarray}\label{Eua-cE}
Z^{{\underline{M}}}= (x^\mu\, ,
\theta^{\check{\underline{\alpha}}})\; , \quad
{\check{\underline{\alpha}}}=1,\ldots , 32\; , \qquad \mu=0,1,\ldots, (D-1) \qquad (D=10,11) \; \qquad
\end{eqnarray}   and supervielbein
form by
\begin{eqnarray}\label{Eua-cE}
{E}^{\underline{A}}:=
dZ^{\underline{M}}E_{\underline{M}}{}^{\underline{A}}(Z)=
 ({E}^{\underline{a}}, {\cal E}^{\underline{\alpha}})\; , \qquad \cases{ {\underline{\alpha}}=1,\ldots ,
32 \; , \cr {\underline{a}}=0,1,\ldots, (D-1)\; } \qquad (D=10,11)\; .   \qquad
\end{eqnarray}
We find convenient, following \cite{IB+PKT=08-2}, to use different symbols the $D$--component bosonic bosonic and for the 32-component fermionic supervielbein forms: ${E}^{\underline{a}}:=
dZ^{\underline{M}}E_{\underline{M}}{}^{\underline{a}}(Z)$ and ${\cal E}^{\underline{\alpha}}:=
dZ^{\underline{M}}{\cal E}_{\underline{M}}{}^{\underline{\alpha}}(Z)$, respectively.

The supervielbein (\ref{Eua-cE}) describes supergravity when it obeys the set of superspace constraints
\cite{CremmerFerrara80,BrinkHowe80,Howe+West=1983,Bellucci+Gates+89} the most essential of which are collected in the expression for the
bosonic torsion two form
\begin{eqnarray}
\label{Ta=11D} & T^{\underline{a}}:= DE^{\underline{a}} =
-i{\cal E}\wedge \Gamma^{\underline{a}} {\cal E} \; ,\qquad {\cal E}\wedge \Gamma^{\underline{a}} {\cal E}:= {\cal E}^{\underline{\alpha}}\wedge
{\Gamma}^{\underline{a}}_{\underline{\alpha}\underline{\beta}}  {\cal E}^{\underline{\beta}}\; . \qquad
\end{eqnarray}
Here and below we write explicitly the exterior product symbol $\wedge$. \footnote{The exterior product of a $q$-form $\Omega_q$ and a $p$-form $\Omega_p$ has the property $\Omega_q \wedge \Omega_p= (-1)^{p q} \Omega_p \wedge \Omega_q $ if at least one of two differential forms is bosonic; when both are fermionic, an additional $(-1)$ multiplier appears in the {\it r.h.s.}. The exterior derivative acts on the products of the forms `from the right': $d(\Omega_q \wedge \Omega_p)=  \Omega_q \wedge d\Omega_p + (-1)^{p} d\Omega_q \wedge \Omega_p$. In particular, $T^{\underline{a}}:= DE^{\underline{a}}= dZ^{\underline{M}}\wedge DE_{\underline{M}}{}^{\underline{a}}(Z)$ so that Eq. (\ref{Ta=11D}) implies $D_{\underline{M}}E_{\underline{N}}{}^{\underline{a}}(Z)- (-1)^{\epsilon({{\underline{M}})} \cdot \epsilon{({\underline{N}}})} D_{\underline{N}}E_{\underline{M}}{}^{\underline{a}}(Z)= +2i (-1)^{\epsilon({{\underline{N}}})}{\cal E}_{\underline{M}}\Gamma^{\underline{a}} {\cal E}_{\underline{N}}$, where $\epsilon ({\underline{M}})$ is the Grassmann parity of $Z^{{\underline{M}}}$, $\; \epsilon({{\underline{a}}})=0$, $\epsilon({{\underline{\alpha}}})=1$.}
In the 11D case ${\Gamma}^{\underline{a}}_{\underline{\alpha}\underline{\beta}}  = ({\Gamma}^{\underline{a}}C)_{\underline{\alpha}\underline{\beta}}= {\Gamma}^{\underline{a}}_{\underline{\beta}\underline{\alpha}}$ where ${\Gamma}^{\underline{a}}= ({\Gamma}^{\underline{a}}){}_{\underline{\alpha}}{}^{\underline{\beta}}$ is the 11D Dirac matrix and $C$ is 11D charge conjugation matrix, which are imaginary in our mostly minus notation \begin{eqnarray}\label{eta=+-...-}
\eta^{\underline{a}\underline{b}}= diag(+,- , \ldots, -) \;  . \qquad
\end{eqnarray}

For $D=10$ type II cases it is convenient to split  the fermionic supervielbein in two 16 component Mojorana-Weyl spinor 1-forms
 \begin{eqnarray}\label{Eual=II}
{\cal E}^{\underline{\alpha}}=\cases{  (E^{\alpha 1}\, , \;
  E_{\alpha}^{2} \; ) \qquad for\; type \; IIA \cr
 (E^{\alpha 1}\, , \,
 E^{\alpha 2}) \qquad for\; type \; IIB } \;   \qquad
\end{eqnarray}
In this notation the mains supergravity constraints (\ref{Ta=11D}) read
\begin{eqnarray}
\label{Ta=IIA} & T^{\underline{a}}:= DE^{\underline{a}} =
-i({E}^{1}\wedge \sigma^{\underline{a}} {E}^{1}+  {E}^{2}\wedge
\tilde{\sigma}^{\underline{a}} {E}^{2})
 \qquad for \; type \;\;  IIA \; ,\qquad
\\ \label{Ta=IIB} & T^{\underline{a}}:= DE^{\underline{a}} = -i
({E}^{1}\wedge
\sigma^{\underline{a}} {E}^{1} + {E}^{2}\wedge \sigma^{\underline{a}} {E}^{2})\qquad  for \;\; type \; IIB \; , \qquad \end{eqnarray}
where  $\sigma^{\underline{a}}:=
\sigma^{\underline{a}}_{\alpha\beta}=\sigma^{\underline{a}}_{\beta\alpha}$
and $\tilde{\sigma}_{\underline{a}}:=
\tilde{\sigma}_{\underline{a}}^{\alpha\beta}=\tilde{\sigma}_{\underline{a}}^{\beta\alpha}$,
are $D=10$ Pauli matrices which obey
\begin{eqnarray}
\label{10DPauli}
\sigma^{\underline{a}}\tilde{\sigma}^{\underline{b}}+\sigma^{\underline{b}}\tilde{\sigma}^{\underline{a}}=2\eta^{\underline{a}\underline{b}} \; = diag(+,-,\ldots ,-) \; , \qquad
 \sigma_{\underline{a}\alpha(\beta} \sigma^{\underline{a}}{}_{\gamma\delta)}\equiv 0 \; , \qquad \tilde{\sigma}{}^{\underline{a}\alpha(\beta} \tilde{\sigma}{}_{\underline{a}}{}^{\gamma\delta)}\equiv 0
\; . \qquad \end{eqnarray}


\section{Superembedding equation as a basis of superembedding approach to D-branes and M-branes}
\setcounter{equation}{0}

Following the so--called STV approach to superparticles and
superstrings \cite{stv,DGHS93}\footnote{ STV abbreviates the family names of
Dmitri Sorokin, Vladimir Tkach and Dmitri Volkov, the authors of \cite{stv}.  These approach to description of
Brink-Schwarz superparticles and Green-Schwarz superstring was also
called 'twistor-like'. See \cite{Dima99} for the review and more
references and \cite{vz,stvz} for related studies of the connection
between Brink--Schwarz and spinning superparticles aimed to relate
spinning (NSR) string and Green--Schwarz superstring already at the
classical level. This line was further continued in \cite{Uvarov}.}
 the superembedding approach
\cite{bpstv,hs96,hs2,bst97,Dima99,B00,IB08:Q7} describes the dynamics of
super-p-brane in terms of embedding of a {\it worldvolume
superspace} into the {\it target superspace}.

\subsection{Worldvolume superspaces $W^{(p+1|16)}$}

The target superspaces of D$p$ branes (M-branes) were described in sec. 1.3.
Their worldvolume superspaces
${\cal W}^{(p+1|16)}$ have $d=p+1$ bosonic and $16$ fermionic
dimensions. We denote the local coordinates of ${\cal W}^{(p+1|16)}$
by
\begin{eqnarray}
\label{zeta=}  \zeta^{{\cal M}}=(\xi^m,\eta^{\check{\alpha}}) \; ,  \qquad
m=0,1,.., p\; , \qquad \check{\alpha}=1,...,16\; .
\end{eqnarray}
The embedding of ${\cal W}^{(p+1|16)}$
into the $D=10$ type II ($D=11$) target superspace $\Sigma^{(10|16+16)}$ ($\Sigma^{(11|32)}$) can be
described in terms of coordinate functions
$\hat{Z}^{{\underline{M}}}(\zeta)=
(\hat{x}{}^{\underline{m}}(\zeta)\, ,
\hat{\theta}^{\check{\underline{\alpha}}}(\zeta))$,
\begin{eqnarray}
\label{WinS}  W^{(p+1|16)}\in \Sigma^{(D|32)} : \hspace{1.5cm}
\fbox{$Z^{\underline{M}}= \hat{Z}^{\underline{M}}(\zeta) $}= (\hat{x}^{\underline{m}}(\zeta)\;
, \hat{\theta}^{\check{\underline{\alpha}}}(\zeta ))\;  , \qquad
\end{eqnarray}
$ D=10, 11$, $\quad\underline{m}=0,1,.., (D-1)$, $\quad\check{\underline{\alpha}}=1,...,32$.

\subsection{The superembedding equation}

A particular beauty of the superembedding approach consists in that,
for all known superbranes, the embedding of the worldvolume
superspace into the target superspace is characterized by a
universal equation which is called the {\it superembedding
equation}. This geometrical equation (the name 'geometrodynamic
equation' was used in \cite{DGHS93}) restricts the coordinate
functions $\hat{Z}^{\underline{M}}(\zeta)$ and, in some cases,
completely determines the dynamics of superbrane.

To write the most general form of this superembedding equation let
us denote the supervielbein of $W^{(p+1|16)}$ by
\begin{eqnarray}
\label{eA=ea+} e^A= d\zeta^{{\cal M}} e_{{\cal M}}{}^{A}(\zeta) =
(e^a\; , \; e^\alpha) \; , \qquad a=0,1,\ldots , p\; , \qquad
\alpha=1,\ldots, 16 \; ,
\end{eqnarray}
and write the general decomposition of the pull--back of the
supervielbein ${E}^{\underline{A}}(Z)$ of target  superspace, Eq. (\ref{Eua-cE}),   to ${\cal W}^{(p+1|16)}$,
$\hat{E}^{\underline{A}}:=E^{\underline{A}}(\hat{Z})$, on this basis,
\begin{eqnarray}
\label{hEa=b+f}
 \hat{E}^{\underline{A}}:= E^{\underline{A}}(\hat{Z})=
d\hat{Z}^{\underline{M}}
E_{\underline{M}}{}^{\underline{A}}(\hat{Z}) = e^b \hat{E}_b^{\,
\underline{A}} + e^\alpha \hat{E}_{\alpha}{}^{\underline{A}} \; .
\qquad
\end{eqnarray}
Notice that  the coincidence of the notation $\alpha, \beta$ for the 10D Majorana--Weyl spinor indices of  the  chiral supervielbein forms  of the target type II superspace ($E^{1,2}$ in (\ref{Eual=II})) and for the indices enumerating the fermionic supervielbein of the worldvolume superspace is not occasional and is acceptable because,  among the $D=10$ objects,  we will discuss D-branes but not fundamental strings (F$1$-branes). We will comment on this more in the next section. In sec. 4 devoted to M-brane  we change the notation and substitute a multiindex $\alpha q$ for $\alpha$ in Eqs. (\ref{eA=ea+}), (\ref{hEa=b+f}).

The superembedding equation states that the bosonic supervielbein
form has zero projection on the worldvolume fermionic supervielbein
form. This is to say, it reads
\begin{eqnarray}
\label{SembEq}
 \fbox{$\hat{E}_{\alpha}{}^{\underline{a}}:= \nabla_{\alpha}
 \hat{Z}^{\underline{M}}\,
E_{\underline{M}}{}^{\underline{a}}(\hat{Z}) =0 \;$}\;  ,  \qquad
\nabla_\alpha:=e_\alpha^{{\cal M}}(\zeta) \partial_{{\cal M}}\, ,
\quad \zeta^{{\cal M}}=(\xi^m  ,  \eta^{\check{\alpha}} )\; .
\end{eqnarray}
It can be also presented in an equivalent form of
\begin{eqnarray}
\label{Ei=0} \hat{E}^{i}:= \hat{E}^{\underline{a}}
u_{\underline{a}}{}^i =0\; , \qquad
\end{eqnarray}
where $u_{\underline{a}} ^{\; i}=u_{\underline{a}} ^{\; i}(\zeta) $
are $(D-p-1)$ spacelike, mutually orthogonal and normalized
$D$-vector superfields,
\begin{eqnarray}\label{uiuj=1-}
u_{\underline{a}} ^{\; i}u^{\underline{a}\; j}= -\delta^{ij}\; .
\qquad
\end{eqnarray}
Eq. (\ref{Ei=0}) means that $u_{\underline{a}}{}^{i}\;$ are orthogonal to the worldvolume
superspace. We can complete their set till a complete {\it moving frame} by
adding $d=(p+1)$ mutually orthogonal and normalized $D$-vector
superfields $u_{\underline{a}} ^{\; b}=u_{\underline{a}}^{\; b}(\zeta)$
which are  tangential to the worldvolume superspace,
\begin{eqnarray}\label{utut=1tt}
u_{\underline{a}}^{\; a} u^{\underline{a}\; i}=0\; , \qquad  u_{\underline{a}} ^{\; a}
\eta^{\underline{a}\underline{b}}u_{\underline{b}} ^{\; b}=\eta^{ab}
\;  , \qquad \cases{ a,b=0,1,\ldots , p\; , \cr \underline{a}\, , \,
\underline{b}=0,1,\ldots , (D-1)\; .} \qquad
\end{eqnarray}
Their contraction with the pull--back $\hat{E}^{\underline{a}}$ of the
target superspace bosonic supervielbein ${E}^{\underline{a}}$
provides us with a set of $d=(p+1)$ linearly independent
nonvanishing one-forms, which can be used as bosonic supervielbein
of the worldvolume superspace,
\begin{eqnarray}\label{Eua=ea}
\hat{E}^a:= \hat{E}^{\underline{b}} u_{\underline{b}} ^{\; a} = e^a
\; .  \qquad
\end{eqnarray}
This $e^a$ is referred to as (super)vielbein form induced by the (super)embedding. Considered together, Eqs. (\ref{Ei=0}) and (\ref{Eua=ea}) imply
\begin{eqnarray}\label{Eua=eua}
\hat{E}^{\underline{a}} = e^b u_b^{\; \underline{\,a}}
\; .  \qquad
\end{eqnarray}
This is one more equivalent form of the superembedding equation.
Indeed, Eqs. (\ref{Eua=ea}) and (\ref{Ei=0}) can be obtained
contracting (\ref{Eua=eua}) with $u_{\underline{a}}{}^a$ and
$u_{\underline{a}}{}^i$ respectively. On the other hand, decomposing
(\ref{Eua=eua}) on the worldvolume supervielbein one arrives at the original form (\ref{SembEq}) of the
superembedding equation. As a by-product on this way one derives the expression for the moving frame vectors $u_{ \underline{\,b}}{}^a(\zeta)$ in terms of the  (linear combination of the)
bosonic derivatives of the coordinate functions,
\begin{eqnarray}\label{ua=EaM2}
u_b{}^{\underline{\, a}} = \hat{E}_b{}^{\underline{\, a}}:= D_b
\hat{Z}^{\underline{M}}(\zeta) E_{\underline{M}}{}^{\!
\underline{\,a}}\, (\hat{Z}(\zeta)) \; . \qquad
\end{eqnarray}

To obtain the consequences of the superembedding equation one can study its integrability (selfconsistency) conditions
\begin{eqnarray}\label{DEi=0=} 0=D\hat{E}^i=
\hat{T}^{\underline{a}} u_{\underline{b}}{}^i + e^b \wedge u_b^{\;
\underline{a}} Du_{\underline{a}}{}^i = - i
{\hat{\cal E}} \wedge \Gamma^{\underline{a}}{\hat{\cal E}} \; u_{\underline{b}}{}^i + e^b \wedge u_b^{\;
\underline{a}} Du_{\underline{a}}{}^i \; .  \qquad
\end{eqnarray}
To this end one has to define the $SO(1,D-1)$ and $SO(D-p-1)$ connection, $\Omega^{bc}=- \Omega^{cb}= d\zeta^M \Omega_M^{bc}$ and $\Omega^{ij}=- \Omega^{ji}= d\zeta^M \Omega_M^{ij}$, entering the $SO(1,D-1)\times SO(1,p)\times SO(D-p-1)$ covariant derivatives
\begin{eqnarray}\label{Dui:=}
Du_{\underline{a}}{}^b:= du_{\underline{a}}{}^b + \omega_{\underline{a}}{}^{\underline{b}} u_{\underline{b}}{}^b +  u_{\underline{a}}{}^c \Omega_c{}^b \qquad and \qquad Du_{\underline{a}}{}^i:= du_{\underline{a}}{}^i + \omega_{\underline{a}}{}^{\underline{b}} u_{\underline{b}}{}^i +  u_{\underline{a}}{}^j \Omega^{ji}
\end{eqnarray}
 acting on the moving frame superfields and superforms in (\ref{DEi=0=}).

\subsection{Moving frame and induced connection on  $W^{(p+1|16)}$ }

Notice that the orthogonality and normalization conditions for the
moving frame vectors $u_{\underline{a}}{}^{b}$ and
$u_{\underline{a}}{}^{j}$ imply that the  $D\times D$ matrix $U$
composed of their components, which we call {\it moving frame
matrix}, is pseudo--orthogonal ($U\eta U^{\!^T}=\eta$),  {\it i.e.}
Lorentz group valued
\begin{eqnarray}\label{UinSO}
U_{\underline{a}}^{(\underline{b})} := \left(u_{\underline{a}} ^{\;
b} \; , \; u_{\underline{a}} ^{\; i} \right) \qquad \in \qquad
SO(1,D-1) \; . \qquad
\end{eqnarray}
These moving frame vectors (also called Lorentz harmonics, see
\cite{Sok} as well as \cite{bpstv,Dima99,B00} and refs. therein) can be used to construct the $SO(1,p)$ and $SO(9-p)$
connections on the worldvolume superspace. In the case of flat
target superspace these would be given by the corresponding Cartan
forms $u^{\underline{c}a} du_{\underline{c}} ^{\; b}$ and
$u^{\underline{a}i} du_{\underline{a}} ^{\; j}$. In the case of
curved target superspace one has to use the pull--back of the spin
connection to make the definition $SO(1,9)$ covariant. It is
convenient to write the definition of the connections implicitly, using the
$SO(1,D-1)\times SO(1,p)\times SO(D-p-1)$ covariant derivatives action
on the moving frame vector, Eqs. (\ref{Dui:=}),
\begin{eqnarray}\label{Dp:Du=Om}
Du_{\underline{b}}{}^a = u_{\underline{b}}{}^i \Omega^{ai}\; ,
\qquad Du_{\underline{b}}{}^i = u_{\underline{b}a} \Omega^{ai}\; .
\qquad
\end{eqnarray}
Both equations in  (\ref{Dp:Du=Om}) involve the 1--form  $\Omega^{ai}$. This generalizes the ${SO(1,9)\over SO(1,p)\otimes
SO(8-p)}$ covariant Cartan form and obeys the generalized Peterson-Codazzi
equations
\begin{eqnarray}
\label{MC=PC} D\Omega^{a i} = \hat{{\bb R}}^{a i}\, , \qquad
\hat{{\bb R}}^{a i} :=(u\hat{{\bb R}}u)^{a i}:= \hat{{\bb
R}}{}^{\underline{c}\underline{b}}{u}_{\underline{c}} ^{\;
{a}}{u}_{\underline{b}} ^{\; {i}} \; , \qquad
\end{eqnarray}
where $\hat{{\bb R}}{}^{\underline{c}\underline{b}}$ is the
pull--back of the curvature of the corresponding type II target superspace. The
curvatures of the induced $SO(1,p)$ and $SO(9-p)$ connections,
$r^{ab}=-r^{ba}$ and ${\bb G}^{ij}= -{\bb G}^{ji}$, are defined, as
usually, by Ricci identities, {\it e.g.} $DDu_{\underline{b}}{}^a=:
\hat{{\bb R}}_{\underline{b}}{}^{\underline{c}}u_{\underline{c}}{}^a
- u_{\underline{b}}{}^b r_b{}^a$, $DDu_{\underline{b}}{}^i=:
\hat{{\bb R}}_{\underline{b}}{}^{\underline{c}}u_{\underline{c}}{}^i
+ u_{\underline{b}}{}^j {\bb G}^{ji}$. Using (\ref{Dp:Du=Om})
and (\ref{MC=PC}) one finds the following generalizations of the
Gauss and Ricci equations (see \cite{bpstv})
\begin{eqnarray}\label{MC=GandR}
r^{ab} = (u{\bb R}u)^{ab} + \Omega^{ai} \wedge \Omega^{bi}\; , \;
\qquad {\bb G}^{ij}=  (u{\bb R}u)^{ij} - \Omega_a{}^{i}\wedge \,
 {\Omega}^{aj}\, \; . \qquad
\end{eqnarray}

Now we can further specify the integrability condition (\ref{DEi=0}) for the superembedding equation (\ref{Ei=0}): \begin{eqnarray}\label{DEi=0} 0=D\hat{E}^i= - i
{\hat{\cal E}} \wedge \Gamma^{\underline{b}}{\hat{\cal E}} \; u_{\underline{b}}{}^i + e_b \wedge
\Omega^{bi} \; .  \qquad
\end{eqnarray}
Decomposing $\Omega^{bi}$ on the worldvolume supervielbein, $\Omega^{bi}=e^\alpha \Omega_\alpha{}^{bi}+ e^b\Omega_a{}^{bi}$ we see that (\ref{DEi=0}) involves only antisymmetric part $\Omega_{[a\,b]}{}^{ i}$ of the bosonic coefficient, while its symmetric part,
\begin{eqnarray}\label{Kabi=M2}
\Omega_{(a\, b)}{}^i= K_{a\, b}{}^i:= -
D_{(a}\hat{E}_{b)}{}^{\!\underline{\,c}}\; u_{\underline{c}}{}^i\; ,
\qquad
\end{eqnarray}
remains free on this stage. The last equality in (\ref{Kabi=M2}) is derived using
Eq. (\ref{ua=EaM2}).  $K_{a\, b}{}^i$ can be recognized as the
(superfield generalization of the) second fundamental form of the
worldvolume superspace considered as a surface in the target
superspace. Then the generalized Cartan form (1--form) gives a superform generalization
of the second fundamental form $K_{ab}{}^i$. \footnote{This is 0-form, but has a natural bosonic one-form representation as $e^aK_{ab}{}^i$. The term 'second fundamental form' does not refer to {\it differential forms}, usually associated with antisymmetric tensors; it is from the language of the classical surface theory where the term 'first fundamental form' refers to the metric. See refs. in \cite{bpstv}. }

To move further we have to impose one more conventional constraint to determine the fermionic supervielbein form of the worldvolume superspace $e^\alpha$. This latter, although excluded from the decomposition of the pull--back of the bosonic supervielbein by the superembedding equation (\ref{SembEq}), does  enter the decomposition of the fermionic supervielbein ${\cal E}^{\underline{\alpha}}= e^\beta V_\beta{}^{\underline{\alpha}} + e^a \psi_a^{\underline{\alpha}}$ which is involved in the selfconsistency condition (\ref{DEi=0=}) and also in the expression for the torsion 2-form of the induced geometry of the worldvolume superspace,
\begin{eqnarray}\label{DEa=ta} De^a = - i
{\hat{\cal E}} \wedge \Gamma^{\underline{b}}{\hat{\cal E}} \; u_{\underline{b}}{}^a \; .  \qquad
\end{eqnarray}

The fermionic supervielbein form $e^\alpha$ of the worldvolume
superspace $W^{(p+1|16)}$ can also be induced by superembedding. At this stage, when studying the case of M-branes and fundamental string, one has to introduce one more notion: the spinor moving frame variables or spinorial Lorentz harmonics \cite{bpstv} (used before in studying superparticles \cite{B90,Ghsds} and twistor-like spinor moving frame action for superstrings and super-p-branes \cite{BZ-p}). These objects which are used to relate the worldvolume superspace fermionic supervielbein with the pull--back of its  target superspace counterpart, $e^\alpha = {\cal E}^{\underline{\beta}} V_{\underline{\beta}}{}^\alpha$, will be discussed in Sec. \ref{M-braneSec} devoted to the superembedding approach to M-branes.

Surprisingly, the case of D$p$-brane happens to be simpler in the sense that one can escape the necessity to introduce the notion of spinor moving frame, at least at this stage. This is why we begin a more concrete part of our review of the superembedding approach from the case of D-branes.

\section{Superembedding approach to D$p$-branes.   }
\setcounter{equation}{0}

The superembedding approach to D$p$--branes was used to describe their dynamics in \cite{hs96} , were it was shown that the superembedding equation was shown to produce their equations of motion some monthes before the generic nonlinear DBI+WZ action was found in \cite{Dpac,Dpac2,B+T=Dpac}\footnote{For the particular case of D$2$--brane  the action had been found in earlier \cite{Townsend95} by applying the d=3 scalar-vector duality   to the M$2$--brane action \cite{BST87}.}. It was further studied in \cite{bst97}, where, in particular, the explicit form of the D$p$-brane fermionic equations was derived for the first time (for the particular case of D$4$-brane these might be extracted from the M5-brane fermionic equations which were presented before in \cite{hs2}).  See   \cite{HS+Chu=PLB98,ABKZ,B00} and references in \cite{IB+DS06,IB08:Q7} for further development.

As it has been already noticed,  basic equation of the superembedding approach to D$p$--brane is the superembedding equation (\ref{SembEq}) equivalent to (\ref{Ei=0}). All the formulae of Sec. 2 are valid for this case so that we will continue by specifying the fermionic supervielbein forms of the D$p$-brane  worldvolume superspace and using it to extract the consequences of the superembedding equation.

\subsection{Fermionic supervielbein induced by superembedding and the first consequence of superembedding equation}

When describing Dp-branes, it is convenient to identify
$e^\alpha$ with the pull--back to $W^{(p+1|16)}$ of, say, the first
of two target space fermionic supervielbein forms
\begin{eqnarray}\label{Dp:ef=Ef1}
e^\alpha = \hat{E}^{\alpha 1}\; . \quad
\end{eqnarray}
Then the general decomposition of the second fermionic supervielbein
form reads
\begin{eqnarray}\label{Dp:Ef2=}
\cases{
 \hat{E}_{\alpha}^{2}=  e^\beta h_{\beta \alpha} + e^a \chi_{a\alpha} \qquad for \; IIA\; case\; , \quad   \cr \hat{E}^{\alpha 2}=  e^\beta h_\beta{}^\alpha + e^a \chi_a^\alpha \qquad for \; IIB\; case \; .}  \quad
\end{eqnarray}
To resume,
\begin{eqnarray}\label{Dp:E32=IIA}
{\hat{\cal E}}^{\underline{\alpha}} =
 (e^{\alpha}\, , \,   e^\beta h_{\beta \alpha} + e^a \chi_{a\alpha}) \qquad for \; IIA\; case \; ,  \quad \\
 \label{Dp:E32=IIB} {\hat{\cal E}}^{\underline{\alpha}} = (e^{\alpha}\, , \,   e^\beta h_\beta{}^\alpha + e^a \chi_a{}^\alpha ) \qquad for \; IIB\; case  \; . \quad
\end{eqnarray}


Now we are ready to find  the first nontrivial
consequence of the superembedding equation.  Looking at the selfconsistency
conditions (\ref{DEi=0}) for superembedding equation (\ref{Ei=0}), we notice that  the
second term does not contribute to the lowest
dimensional (dim 2, i.e. $\propto e^\beta \wedge e^\alpha$)
component of this differential form equation. Thus, substituting (\ref{Dp:E32=IIA}) or (\ref{Dp:E32=IIB}) into Eq.  (\ref{DEi=0}) we find
\begin{eqnarray}\label{hsih=-si}
 \, h\tilde{\sigma}^{\underline{b}}h^T u_{\underline{b}}{}^i
= -
 \sigma^{\underline{b}} u_{\underline{b}}{}^i
 \qquad for \; type\; IIA\; ,  \qquad \\
\label{hsih=-siIIB}
 \, h{\sigma}^{\underline{b}}h^T u_{\underline{b}}{}^i
= -
 \sigma^{\underline{b}} u_{\underline{b}}{}^i
 \qquad for \; type\; IIB\;  . \qquad
\end{eqnarray}
We can continue by studying the higher dimensional components of
Eq.  (\ref{DEi=0}) and also of the (conventional) equations for the fermionic supervielbein (\ref{Dp:Ef2=}). On this way one finds, in particular, that the field strength $F_{ab}$ of the worldvolume gauge field is related to the spin-tensor $h$ in the decomposition (\ref{Dp:Ef2=}).
However, it is technically  much simpler, using the knowledge on the very existence of the worldvolume gauge field,  to introduce its superform counterpart on the worldvolume superspace, to restrict it by a suitable set of constraints and study their selfconsistency conditions.

\subsection{Constraints for the worldvolume gauge field. }

The constraints for the worldvolume gauge (super)field strength of the $Dp$-brane can be written as
\begin{eqnarray}\label{Dp:F2=dA-B2=}
F_2:= dA -\hat{B}_2 = {1\over 2} e^b\wedge e^a F_{ab} \; ,  \qquad
\end{eqnarray}
where $\hat{B}_2$ is the pull--back  to the worldvolume superspace
$W^{(p+1|16)}$  of the type IIB NS-NS superform potential $B_2$. The
field strength of this is restricted by the constraints which can be collected in the following differential form expressions:
\begin{eqnarray}
\label{H3=IIA} H_{3}:=dB_2 &=& - i {E}^{\underline{a}}\wedge
({E}^{1}\wedge \sigma_{\underline{a}} {E}^{1} -  {E}^{2}\wedge
\tilde{\sigma}_{\underline{a}} {E}^{2}) + \qquad \nonumber \\ && {}\qquad  + {1\over 3!}
{E}^{\underline{c}_3}\wedge{E}^{\underline{c}_2}\wedge
{E}^{\underline{c}_1}
H_{\underline{c}_1\underline{c}_2\underline{c}_3} \; , \qquad for\; type\; IIA \; ,  \\
\label{H3=IIA}  H_{3}:=dB_2 &=& - i {E}^{\underline{a}}\wedge
({E}^{1}\wedge \sigma_{\underline{a}} {E}^{1} -  {E}^{2}\wedge
{\sigma}_{\underline{a}} {E}^{2}) + \qquad \nonumber \\ && {}\qquad + {1\over 3!}
{E}^{\underline{c}_3}\wedge{E}^{\underline{c}_2}\wedge
{E}^{\underline{c}_1}
H_{\underline{c}_1\underline{c}_2\underline{c}_3}\; , \qquad for\; type\; IIB \; .
\end{eqnarray}

The lowest dimensional of the nontrivial  components of the Bianchi identity
\begin{eqnarray}\label{Dp:dF2=-H3}
dF_2= -\hat{H}_3 \;
\end{eqnarray}
is $\propto e^\gamma\wedge e^\beta
\wedge e^a$, this is to say of {\it dim} 2. It implies

 implies
\begin{eqnarray}\label{hsah=ska}
\matrix{ \, h\sigma^{\underline{b}}h^T u_{\underline{b}}{}^a =
\sigma^{\underline{b}} u_{\underline{b}}{}^c k_c{}^a   \qquad for \; IIB \; , \qquad \cr
\, h\tilde{\sigma}^{\underline{b}}h^T u_{\underline{b}}{}^a =
 \sigma^{\underline{b}} u_{\underline{b}}{}^c k_c{}^a  \qquad for\; IIA \; ,\qquad}
 k_a{}^b:=(\eta +F)_{ac}(\eta-F)^{-1}{}^{cb} 
 \; . \qquad
\end{eqnarray}
Notice that this equation relates the spin-tensor $h$, appearing in the decomposition of the pull--back of the fermionic supervielbein form (\ref{Dp:Ef2=}), and the bosonic gauge field strength tensor superfield $F_{ab}=-F_{ba}$.
One can easily check that the matrix $k$, constructed from $F_{ab}$ as in (\ref{hsah=ska}), is SO(1,p) group valued,
{\it i.e.} it obeys $k\eta k^T=\eta$ \cite{ABKZ,IB+DS06},
 \begin{eqnarray}\label{kinSOp}
 k =(\eta +F)(\eta-F)^{-1}\qquad \in \qquad  SO(1,9)\; . \qquad
\end{eqnarray}

Further study shows that the system of superembedding equation plus the worldvolume gauge field constraints (\ref{Dp:F2=dA-B2=}) always contains the dynamical equations among their consequences (and for $p\leq 5$ D$p$-branes \cite{HS+Chu=PLB98} the superembedding equation along suffice for this purposes). However, the details of derivation are $p$-dependent. As an example, below we will give some details for the case of D$0$--brane which will be then used in Sec. \ref{multipleD0}. But before let us discuss a toy example:  D$(-1)$-brane or D-instanton. What one can obtain form the superembedding approach in this case?

\subsection{A toy example: D-instanton (D$(-1)$-brane)}

For instanton the dimension of the bosonic body of the worldvolume superspace is zero, $d=p+1=0$, so that this superspace is purely fermionic $W^{(0|16)}$. Its co-tangent superspace basis contains the fermionic supervielbein $e^\alpha$ only, all the spacetime directions are orthogonal to the
worldvolume superspace so that the moving frame matrix is not
needed. Hence the superembedding equation for D-instanton reads
\begin{eqnarray}\label{Din:Eua=0}
\hat{E}^{\underline{b}} = 0
\; .  \qquad
\end{eqnarray}
The fermionic supervielbein of the worldvolume superspace $e^\alpha$ can be identified with the pull--back $\hat{E}^{\alpha 1}$ of ${E}^{\alpha 1}$, and the general decomposition of the pull--back $\hat{E}^{\alpha 2}$ of ${E}^{\alpha 2}$ reads $\hat{E}^{\alpha 2}=e^\beta h_\beta{}^\alpha$.
The selfconsistency conditions for the superembedding equation implies vanishing of the pull--back
of the target space bosonic torsion,
$0= \hat{T}^{\underline{a}} = D\hat{E}^{\underline{b}}= -i e^\alpha\wedge e^\beta
\, (\sigma^{\underline{a}} + h\sigma^{\underline{a}}h^T )_{\alpha\beta}$.
This results in equation
 \begin{eqnarray}\label{Din:hsh=} h\sigma^{\underline{a}}h^T=- \sigma^{\underline{a}}
\;   \qquad
\end{eqnarray}
which {\it does not} have solution in the case of real $h$. However,
there is an imaginary  solution,
\begin{eqnarray}\label{Din:h=iI} h_\alpha{}^\beta = i \delta_\alpha{}^\beta
\; .  \qquad
\end{eqnarray}
It implies that $\hat{E}^{\alpha 2}= i\hat{E}^{\alpha 1}$ and hence, as far as
\begin{eqnarray}\label{Din:Ef=c}
\hat{E}^{\alpha 1}= - i\hat{E}^{\alpha 2} = e^\alpha \; , \qquad 
 \hat{E}^{\alpha 1}+ i\hat{E}^{\alpha 2} = 0\; , \qquad
\end{eqnarray}
that  both tangent superspace and worldvolume superspace fermionic supervielbeins are complex.
This is in agreement with the well known fact that D-instanton implies Wick rotation {\it i.e.}
exists only in the Euclidean version of the type IIB theory, where the real 16 component Weyl spinor is inevitably complex (versus the existence of real Majorana-Weyl spinor in the case of Lorentz 1+9 signature).

This seems to be the only result one can get from superembedding description of D-instanton. It
is not surprising as far as D-instanton has no dynamics: it is frozen to a point of Euclidean spacetime
(which is expressed by the statement that it is $(-1)$-brane).

\subsection{D$0$-brane in superembedding approach}

In the case of D$0$--brane, this is to say D-particle, there are nine space-like directions orthogonal to the worldline and the tangent to the worldline gives a time-like directions, so that the corresponding set of moving frame vectors $(u_{\underline{a}}^0\, ,
u_{\underline{a}} ^{\; i})$ obeys
\begin{eqnarray}\label{D0:uu=1}
u_{\underline{a}} ^{\; 0}u^{\underline{a}0}=1\; , \qquad
u_{\underline{a}} ^{\; i}u^{\underline{a}0}=0\; . \qquad
u_{\underline{a}} ^{\; i}u^{\underline{a}j}=-\delta^{ij}\; . \qquad
\end{eqnarray}

The worldvolume superspace $W^{(1|16)}$ has only one bosonic direction, $e^a\mapsto e^0$ and
the superembedding equation (\ref{Eua=eua}) (equivalent to (\ref{SembEq})) reads
\begin{eqnarray}\label{Eua=e0u0a}
\hat{E}^{\underline{a}} = e^0 u_0^{\, \underline{\,a}}
\; .  \qquad
\end{eqnarray}
The expressions (\ref{Dp:ef=Ef1}) for the pull--backs of the fermionic supervielbein form  simplifies to  \begin{eqnarray}\label{D0:ef=Ef1}
\hat{E}^{\alpha 1}&=&  e^{\alpha}\; , \qquad \\  \label{D0:Ef2=}
 \hat{E}_{\alpha}{}^{2}&=&  e^\beta h_{\beta \alpha} + e^0\chi_{\alpha} \; . \qquad
\end{eqnarray}
It is convenient to write  the selfconsistency conditions (\ref{hsih=-si}) for the superembedding equation (\ref{Eua=e0u0a}) in the form of
 \begin{eqnarray}\label{D0:hsih=}
 h{\tilde{\sigma}}{}^ih^T  =- {\sigma}^i
\;  \qquad
\end{eqnarray}
 using  the simplified notation
\begin{eqnarray}\label{D0:sui=si}  {\sigma}^0_{\alpha
\beta}:= {\sigma}^{\underline{b}}_{\alpha
\beta}u_{\underline{b}}{}^0 \; , \qquad {\sigma}^i_{\alpha
\beta}:= {\sigma}^{\underline{b}}_{\alpha
\beta}u_{\underline{b}}{}^i \; . \qquad
\end{eqnarray}
These are suggestive as far as the matrices (\ref{D0:sui=si}) and
$\tilde{\sigma}^0_{\alpha
\beta}:= \tilde{\sigma}{}^{\underline{b}}_{\alpha
\beta}u_{\underline{b}}{}^0 $, $\tilde{\sigma}^i_{\alpha
\beta}:= \tilde{\sigma}{}^{\underline{b}}_{\alpha
\beta}u_{\underline{b}}{}^i $ do possess the algebraic properties of D=10 Pauli matrices. However,
one should keep in mind that  they are not constant matrices but rather obey
\begin{eqnarray}\label{D0:Dsui=}
D{\sigma}_{\!_{\alpha \beta}}^0= {\sigma}_{\!_{\alpha
\beta}}^i\Omega^i\; , \qquad D{\sigma}_{\!_{\alpha \beta}}^i=
{\sigma}_{\!_{\alpha \beta}}^0\Omega^i\; , \qquad
\end{eqnarray}
where $\Omega^i$ is the generalized Cartan form defined in (\ref{Dp:Du=Om}). In this notation the
the general solution of Eq.(\ref{D0:hsih=}) reads
\begin{eqnarray}\label{D0:h=} h_{\alpha \beta} = {\sigma}^{0}_{\alpha \beta}\; .
\qquad
\end{eqnarray}

This is the place to comment on  the worldvolume gauge field constraints for the
D$0$-brane  case (worldline gauge field).
For the $p=0$   the {\it r.h.s.} of Eq. (\ref{Dp:F2=dA-B2=}) clearly vanishes so
that the constraints read \begin{eqnarray}\label{D0:F2=0}F_2:=dA-\hat{B}_2=0\;  \qquad
\end{eqnarray} and the Bianchi
identities (\ref{Dp:dF2=-H3}) simplify to $\hat{H}_3=0$. Their  only
nontrivial consequence reads
\begin{eqnarray}\label{D0:hsu0h=}
 h\tilde{\sigma}^{0}h^T  =
 {\sigma}^0\;   \qquad
\end{eqnarray}
Eq. (\ref{D0:hsu0h=}) is satisfied identically by the general solution (\ref{D0:h=}) of Eq. (\ref{D0:hsih=}). This shows  that the gauge field constraints in the case of D$0$-brane are dependent, which is in agreement with the known statement that the superembedding equation alone is sufficient to describe dynamics in this case. On the other hand,  to arrive at the equations of motion
in a simpler way, it is convenient to impose the gauge field
constraints (\ref{D0:F2=0}) on the field strength of the
worldvolume gauge field. Indeed, it is evident without any calculation that the general solution of Eqs. (\ref{D0:hsih=}) and (\ref{D0:hsu0h=}) is given by (\ref{D0:h=}).

Another consequence of the selfconsistency conditions for the superembedding equation
(\ref{Eua=e0u0a}) is that the generalized Cartan form  $\Omega^i$ in (\ref{D0:Dsui=}) is expressed by
\begin{eqnarray}\label{D0:Omi=}
\Omega^{i}= e^0\, K^i - 2i e^\beta
 ({\sigma}^0\tilde{\sigma}^i\chi)_\beta\;  \qquad
\end{eqnarray}
in terms of fermionic superfield
$\chi_\alpha=\hat{E}_{0\alpha}{}^{2}$ and bosonic superfield
\begin{eqnarray}\label{D0:Ki:=}
 K^i&:= - u^i_{\underline{a}}D_0 \hat{E}_0^{\underline{a}}
\; , \qquad \hat{E}_0^{\underline{a}}:=
\nabla_0\hat{Z}^{\underline{M}}{E}_{\underline{M}}{}^{\underline{a}}(\hat{Z})
\; . \qquad
\end{eqnarray}
This latter is the superfield generalization of the mean  curvatures of the particle
worldline in target space. The generalized Cartan form (\ref{D0:Omi=}) gives the superform generalization of this mean curvature for the case of D$0$-brane in type IIA superspace.  It contains $K^i$ as a dim 1 and the fermionic $\chi_{\alpha}=\hat{E}{}_{0\alpha}{}^2$ superfield as a dim 1/2 component; in this sense $\chi_{\alpha}$ is the superpartner of $K^i$.
The bosonic and fermionic equations, which can be now obtained from the
selfconsistency condition for the fermionic Eq. (\ref{D0:Ef2=}), are
formulated in terms of these superfields.

In flat target superspace
the equations of motion imply vanishing of both $\chi_\alpha$ and
$K^i$,
\begin{eqnarray}\label{D0:flatEqm=} \chi_\alpha :=
\hat{E}_{_0}{}^2_{\alpha} =0 \; , \qquad
 K^i:= - u^i_{\underline{a}}D_{_0} \hat{E}_{_0}{}^{\underline{a}} =  0  \; .  \qquad
\end{eqnarray}

In general type IIA supergravity background the fermionic equations
of motion acquires the {\it r.h.s.}
\begin{eqnarray}\label{D0:DiracEq}
\chi_\alpha:= \hat{E}_{_0}{}^2_{\alpha} = \Lambda_\alpha  \;  \qquad
\end{eqnarray}
 defined by
\begin{eqnarray}\label{D0:L=L1+sL2}
 \Lambda_\alpha:=  (\hat{\Lambda}_1-\hat{\Lambda}_2\sigma^0)_\alpha \;  ,
\qquad
\end{eqnarray}
where $\hat{\Lambda}_{1 \alpha}$ and $\hat{\Lambda}_2{}^\beta$ are the  pull--backs of the Grassmann derivatives of the dilaton
superfield,
\begin{eqnarray}\label{D0:Lambda1=}
\Lambda_{\alpha 1} := {i\over 2}{(D_{\alpha 1}{\Phi})}\;  ,
\quad \Lambda_2^{\alpha}\; := {i\over 2}\,{(D_2^{\alpha}\,{\Phi})}\;
. \qquad
\end{eqnarray}
are the  pull--backs of the Grassmann derivatives of the dilaton
superfield. The origin of the {\it r.h.s.} in Eq. (\ref{D0:DiracEq}) is nonvanishing fermionic torsion of the target type IIA superspace \cite{Bellucci+Gates+89}
\begin{eqnarray}\label{Tal1-2=IIA}
& T^{\alpha 1} =
 - 2i E^{\alpha 1}\wedge E^{\beta 1} \Lambda_{\beta 1} + i E^{1}\sigma^{\underline{a}}\wedge E^{1}\,
\tilde{\sigma}_{\underline{a}}^{\alpha\beta} \Lambda_{\beta 1} +
\propto E^{\underline{b}}  \, ,  \, \qquad \nonumber \\
 & T_{\alpha}^{ 2} =
 -   2i  E^2_{\alpha}\,\wedge
E^2_{\beta} \; \Lambda_2^{\beta}\,  + i
E^{2}\tilde{\sigma}_{\underline{a}}\wedge E^{2}\,
{\sigma}^{\underline{a}}_{\alpha\beta} \, \Lambda_2^{\beta } +
\propto E^{\underline{b}}   \; . \qquad
\end{eqnarray}
The bosonic equation for D$0$-brane in general supergravity
background reads
\begin{eqnarray}\label{D0:bEqm=}
 K^i&:= - u^i_{\underline{a}}D_0 \hat{E}_0^{\underline{a}} = {1\over 16}
 \tilde{\sigma}{}^{i\alpha\beta} (t_{\alpha\beta} - D_{\alpha} \Lambda_{\beta} )
 + {7i\over 8}(\hat{\Lambda}_2\sigma^{0i}\hat{\Lambda}_1) = \qquad \nonumber \\
 & =e^{\hat{\Phi}}\hat{R}{}^{0i}+\widehat{D^i\Phi}+ {\cal O}(fermi^2)\;  , \qquad
\end{eqnarray}
 where
 \begin{eqnarray}\label{D0:tab=}t_{\alpha\beta}= \left( \widehat{T}_{\alpha 1\,
\underline{a}}{}^2_{\beta }+ \sigma^0_{ \alpha\gamma}
\widehat{T}^{\gamma}_{2}{}_{\underline{a}}{}^2_{\beta}-
\widehat{T}_{\alpha 1 \underline{a}}{}^{\delta 1} \sigma^0_{\delta
\beta }- \sigma^0_{ \alpha\gamma}
\widehat{T}^{\gamma}_{2}{}_{\underline{a}}{}^{\delta 1}
\sigma^0_{\delta \beta }\right)\, u^{\underline{a}\, 0}\;  . \qquad
\end{eqnarray} To arrive
at the second line of Eq. (\ref{D0:bEqm=}), written explicitly up to
the fermionic contributions, one has to use the explicit form of the dimension $1$ target space torsion spin-tensors, entering (\ref{D0:tab=}), and of the derivatives of fermionic superfield $D_{\alpha} \Lambda_{\beta}$ which  can be found in Appendix B.

\section{M-branes in the superembedding approach
\label{M-braneSec}}
\setcounter{equation}{0}

The basic superembedding equation describing the dynamics of M2- and M5-branes have the same form (\ref{SembEq}), or equivalently (\ref{Ei=0}). However, in these cases the fermionic supervielbein ${\cal E}{}^{\underline{\alpha}}$ is in the minimal $32$--component D=11 Majorana spinor representation so that the trick we used in the case of Dp-branes does not work and the relation between $\hat{{\cal E}}{}^{\underline{\alpha}}$ and the worldvolume superspace fermionic supervielbein form $e^{\alpha q}$  is now more complicated.

Notice that, when studying 11D M-branes (and also fundamental strings in D=10) it is convenient to denote the fermionic supervielbein of the worldvolume superspace $W^{(p+1|32)}$ by $e^{\alpha q}$,
\begin{eqnarray}\label{ea->eaq}
e^{\alpha}{}_{\; of\; secs. \; 2 , 3\; and \; 5}\quad \longleftrightarrow  \quad e^{\alpha q}{}_{\; of\; this\; sec.} \quad with \qquad \cases{ \alpha = 1,\ldots , s_p\; , \cr ^{_{s_p:= dim(Spin(1,p))}}\; , \cr q=1,\ldots , {16\over s_p}  \; ,}   \qquad
\end{eqnarray}
{\it i.e.} to split the 16-valued (multi)index of this fermionic one-form on the $Spin(1,p)$ index $\alpha$ ($\alpha=1,2$ for M2- and $\alpha=1,2,3,4$ for M5--brane) and the $Spin(D-p-1)$ index $q$ ($q=1,\ldots, 8$ form M2- and $q=1,\ldots, 4$ form M5-brane).

The fermionic supervielbein $e^{\beta p}$ induced by superembedding can be defined in terms of the pull--back $\hat{\cal E}{}^{\underline{\alpha}}$ of the $D=11$ targets superspace fermionic supervielbein ${\cal E}{}^{\underline{\alpha}}$ with the use of
$16\times 32$ matrix  $v_{\underline{\alpha}}{}^{\beta p}$ of rank 16,
\begin{eqnarray}\label{e=EV}
e^{\beta p}= \hat{\cal E}{}^{\underline{\alpha}}v_{\underline{\alpha}}{}^{\beta p}\; .  \qquad
\end{eqnarray}
The simplest choice of $v_{\underline{\alpha}}{}^{\beta p}$ to be a $32\times 16$ block of unity matrix clearly breaks $SO(1,10)$ Lorentz symmetry (at least down to $SO(1,9)$ in which case we arrive at equation equivalent to (\ref{Dp:ef=Ef1})). To preserve the 11D Lorentz symmetry we have to assume that   $v_{\underline{\alpha}}{}^{\beta p}$ is a $32\times 16$ matrix superfield. It is convenient to consider it as a $32\times 16$ block of an $Spin(1,10)$ group valued  $32\times 32$ matrix superfield
\begin{eqnarray}\label{VharmM2} V^{~(\underline{\alpha})}_{\underline{\beta}}= \left(
v^{\;\;\alpha q}_{\underline{\beta}},~ v_{\underline{\beta}\, \alpha \dot q} \right)\;
\in \; Spin(1,10) \; , \qquad \cases{ \alpha ,\beta =1,2 \; , \cr q=1,
\ldots , 8\; ,} \quad for \;\; M2-brane \; , \quad
\\ \label{VharmM5} V^{~(\underline{\alpha})}_{\underline{\beta}}= \left(
v^{\;\;\alpha q}_{\underline{\beta}},~ v_{\underline{\beta}}{}_{\alpha}^{\; q} \right)\;
\in \; Spin(1,10) \; , \quad \cases{ \alpha ,\beta =1,2,3,4 \; , \cr q=1,2,3,4
\; , } \quad for \;\; M5-brane \; . \quad
\end{eqnarray}
These {\it spinor moving frame superfields} (also called spinor Lorentz harmonics \cite{B90,Ghsds,BZ-p}) describe the spinor representation of the same $SO(1,10)$ Lorentz rotation the vector representation of which is described by the moving frame variables (\ref{UinSO}) and, hance, carry the same local  degrees of freedom as the moving frame vectors\footnote{These moving frame vectors can be identified with derivatives of the coordinate functions (see Eq. (\ref{ua=EaM2})) so that one can either state that they are auxiliary fields which do not bring new {\it dynamical} degrees of freedom, or, equivalently, say that they carry some `momentum' part of degrees of freedom; in other words, they are counterparts of momentum variable $p$ in the first order formulation of the particle mechanics  $S=\int d\tau p\dot{q}- \int d\tau e(p^2-m^2)/2$. }.

The $Spin$ group, the double covering of the Lorentz group $SO$, is defined by the
conditions of the preservation of the gamma matrices. Hence the above mentioned relation between vector
and spinor moving frame variables (vector and spinor Lorentz harmonics) of Eqs.
(\ref{UinSO}) and (\ref{VharmM2}) (or (\ref{VharmM5})) is given by
\begin{eqnarray}\label{VGVT=uHa}
V\Gamma^{(\underline{a})}V^T = \Gamma^{\underline{b}}
U_{\underline{b}}{}^{(\underline{a})} \qquad \Rightarrow \qquad \cases{V\Gamma^{a}V^T =
\Gamma^{\underline{b}} u_{\underline{b}}{}^{a}\; , \cr V\Gamma^{i}V^T =
\Gamma^{\underline{b}} u_{\underline{b}}{}^{i}}
 \; ,  \qquad
\end{eqnarray}
or, equivalently, by
\begin{eqnarray}\label{VTGV=uHa}
V^T\tilde{\Gamma}^{\underline{a}}V = \tilde{\Gamma}^{(\underline{b})}
U_{(\underline{b})}{}^{\underline{a}}= \tilde{\Gamma}^{b} u_{b}{}^{\underline{a}} - \tilde{\Gamma}^{i}
u^{i\underline{a}}\; .   \qquad
\end{eqnarray}
In the dimensions where the charge conjugation matrix $C$ exists, including the
cases of $D=11$ we are interested in here (but not in D=10 ${\cal N}=1$ and type IIB cases), the
condition of its conservation should be also listed among the defining relations of the
spinorial moving frame variables,
\begin{eqnarray}\label{VTCV=C}
VCV^T = C\; ,    \qquad V^TC^{-1}V = C^{-1}\; .    \qquad
\end{eqnarray}
These relations imply that the inverse spinor moving frame matrix $V^{-1}$,
\begin{eqnarray}\label{V-1:=M2}
V^{-1}{}_{(\underline{\alpha})}{}^{\underline{\beta}} \equiv V_{(\underline{\alpha})}{}^{\underline{\beta}} := \left(
iv_{{\alpha q}}{}^{\underline{\beta}} \, , \, iv_{\dot{q}}^{{{\alpha}}}{}{}^{\underline{\beta}}
\right)\quad for\; M2-brane \; , \qquad
 \\
 \label{V-1:=M5}
V^{-1}{}_{(\underline{\alpha})}{}^{\underline{\beta}} \equiv V_{(\underline{\alpha})}{}^{\underline{\beta}} := \left(
v_{{\alpha q}}{}^{\underline{\beta}} \, , \, v^{{{\alpha}}}_{{q}}{}^{\underline{\beta}}
\right)\qquad for\; M5-brane \; ,  \qquad
\end{eqnarray}
obeying
\begin{eqnarray}\label{V-1:=} V_{(\underline{\alpha})}^{\;\;\; \underline{\gamma}} V_{\underline{\gamma}}^{(\underline{\beta})}
=\delta_{(\underline{\alpha})}^{\;\; (\underline{\beta})} = \cases{ diag( \delta_\alpha{}^\beta\delta_q{}^p \, , \delta^\alpha{}_\beta\delta_{\dot{q}}{}^{\dot{p}})\quad for\; M2-brane \cr  diag( \delta_\alpha{}^\beta\delta_q{}^p \, , \delta^\alpha{}_\beta\delta_{{q}}{}^{{p}})\quad for\; M5-brane }
\end{eqnarray}
can be explicitly constructed from the original harmonic matrix (\ref{VharmM2}) or (\ref{VharmM5}),
$V^{-1}=CV^TC^{-1}$. In the case of M2 and M5 brane the  components of the inverse matrices (\ref{V-1:=M2}) and (\ref{V-1:=M5}) are defined by
\begin{eqnarray}\label{v-1=CvM2}
v_{\alpha q}{}^{\underline{\alpha}} = C^{\underline{\alpha}\underline{\delta}}\epsilon_{\alpha\beta}
v_{\underline{\delta}}{}^{\beta q} \; , \qquad  v_{\dot{q}}^{\alpha}{}^{\underline{\alpha}}=  C^{\underline{\alpha}\underline{\delta}}\epsilon^{\alpha\beta}
v_{\underline{\delta}\, \beta \dot{q}}
\qquad for\; M2-brane \; , \qquad
\\
 \label{v-1=CvM5}
v_{\alpha q}{}^{\underline{\alpha}} = iC^{\underline{\alpha}\underline{\delta}}C_{qp}
v_{\underline{\delta}\,\beta}^{p} \; , \qquad  v_{{q}}^{\alpha \underline{\alpha}}=  iC^{\underline{\alpha}\underline{\delta}}C_{qp}  v_{\underline{\delta}}^{\alpha  p}
\qquad for\; M5-brane \; ,  \qquad
\end{eqnarray}
where $C^{\underline{\alpha}\underline{\delta}}$ and $C_{qp}$ are the $D=11=1+10$ and $d=5=5+0$ charge conjugation matrices;  see Appendix A for more details on our notation.  Notice that we found more convenient to introduce $i=\sqrt{-1}$ in the definition of the inverse moving frame matrix components (\ref{VharmM2}) for the case of M2-brane, while in the case of M5-brane we introduced it in the relation between  the components of the inverse and the original moving frame matrices (\ref{v-1=CvM5}). The latter choice looks more natural while the former is explained by that in the case of $p=2$  there exists the $SL(2,{\bb R})=Spin(1,2)$ invariant antisymmetric tensor
$\epsilon^{\alpha\beta}= i\sigma^2$ and its inverse $\epsilon_{\alpha\beta}=- i\sigma^2$ which can be used to rise and to lower the $SL(2,{\bb R})$ ($SO(1,2)$ spinorial) indices; then the use of notation similar to the one accepted for M5-brane case might produce a confusion.

When the charge conjugation matrix does not exist (like in the
$D=10$ $N=1$ case involving the Majorana--Weyl spinor representation) the inverse spinor moving frame variables  are defined just by the constraint $V^{-1}V=I$ (Eq. (\ref{V-1:=})), {\it
i.e.} its dependence on the original harmonics remains implicit.

As the spinor moving frame variables (spinor harmonics) (\ref{VharmM2}) (or (\ref{VharmM5})) carry the same local degrees of freedom as the vector harmonics (moving frame variables) (\ref{UinSO}), their derivatives are expressed through the same generalized Cartan forms (\ref{Dp:Du=Om}).
To find this one just  notice that the Lorentz group $SO(1,D-1)$ and its doubly covered $Spin(1,D-1)$ are locally isomorphic. Then isomorphic are the co-tangent and tangent space to these groups, $spin(1,D-1)\approx so(1,D-1)$. In the case of $SO(1,D-1)$, the latter has the natural basis described by the generalized Cartan
forms $\Omega^{(\underline{a})(\underline{b})}=
u_{\underline{c}}{}^{(\underline{a})}D^Lu^{\underline{c}(\underline{b})}$, where $D^L$ is Lorentz covariant derivative constructed with the use of target superspace spin connection, $D^Lu_{\underline{a}}^{(\underline{b})}= du_{\underline{c}}^{(\underline{b})} + \omega_{\underline{a}}{}^{\underline{c}}u_{\underline{c}}^{(\underline{b})}$.
The isomorphism of $spin(1,D-1)$ and $so(1,D-1)$ algebras is described by the following
universal ($D$--independent) relation between the generalized Cartan forms of $Spin(1,D-1)$ and of $SO(1,D-1)$
\begin{eqnarray}\label{CFspin=CFso}
V^{-1}D^LV &=& {1\over 4} \Omega^{(\underline{a})(\underline{b})}
\Gamma_{(\underline{a})(\underline{b})} := {1\over 4}
(U^{-1}D^LU)^{(\underline{a})(\underline{b})} \Gamma_{(\underline{a})(\underline{b})}
\qquad
\nonumber \\
 &=& {1\over 4} \Omega^{ab}
\Gamma_{ab} + {1\over 4} \Omega^{ij} \Gamma^{ij} - {1\over 2} \Omega^{ai}
\Gamma_a\Gamma^i \; ,   \qquad
\end{eqnarray}
where
$D^LV = dV - {1\over 4} \omega^{\underline{a}\underline{b}}\,
\Gamma_{\underline{a}\underline{b}}V\,$.

In superembedding approach it is convenient to consider the spinor moving frame variables as homogeneous coordinates of the coset ${Spin(1,D-1)\over Spin(1,p)\otimes Spin(D-p-1)}$, using the natural
$Spin(1,p)\otimes Spin(D-p-1)$ gauge symmetry of the embedding of the worldvolume superspace as an identification relation. In practical terms this implies that it is convenient to rewrite Eq. (\ref{CFspin=CFso}) in terms of the ${Spin(1,D-1)\otimes Spin(1,p)\otimes Spin(D-p-1)}$-covariant derivative $D$:
\begin{eqnarray}\label{DV=VOm}
DV &:=& dV - {1\over 4} \omega^{\underline{a}\underline{b}}\,
\Gamma_{\underline{a}\underline{b}}V\, - {1\over 4} V\Gamma_{ab}\, \Omega^{ab} -
{1\over 4} V\Gamma^{ij} \, \Omega^{ij} = - {1\over 2} \Omega^{ai} V\Gamma_a\Gamma^i \; .
 \qquad
\end{eqnarray}

To specify further the above equations one needs to use explicitly an $SO(1,p) \times
SO(D-p-1)$ invariant representation for the $\Gamma$--matrices
\begin{equation}\label{3.13}
\Gamma^{(\underline{a})} = ( \Gamma^{a}, \Gamma^{i} )\;
\end{equation}
so that the further detail are $p$--dependent and will be discussed in the case-by-case manner. The representation convenient for the study of M2- and M5-branes and useful relations for corresponding spinor moving frame variables can be found in Appendix A.

To conclude the general description of the spinor moving frame variables, let us notice that their use is also inevitable when constructing superembedding approach to fundamental string \cite{bpstv} (see \cite{Bandos:2008ba} for recent review and elaboration of a specific case of type IIB superstring in $AdS_5\otimes S^5$ background).

\subsection{Superembedding description of M2--brane (also known as D=11
supermembrane)} \label{M2SEmb}

In this section we will show how the dynamical M2--brane equations follow from the superembedding
equation (\ref{Ei=0}) (equivalent to (\ref{SembEq})) \cite{bpstv},
\begin{eqnarray}\label{Ei=0M2}
\hat{E}^{i}:= \hat{E}^{\underline{\, a}}u_{\underline{\, a}}{}^i = 0
\; .  \qquad
\end{eqnarray}
We have tried to make this section `closed' so that it can be read independently; this explains some repetitions of the statement of the previous sections.

The geometry of the worldvolume superspace is induced by superembedding. This implies, in particular, that its bosonic supervielbein form and $SO(1,2)\otimes SO(8)$ connection are  defined by (\ref{Eua=ea}) and (\ref{Dp:Du=Om}). The fermionic supervielbein of the M2-brane worldvolume superspace ${\cal W}^{(3|16)}$ can be
identified with, say, $\hat{E}^\alpha_q= \hat{\cal E}^{\underline{\beta}}
v_{\underline{\beta}q}{}^\alpha$.
Then,
\begin{eqnarray}\label{Ef2=M20}
\hat{E}^\alpha_q:=\hat{\cal E}^{\underline{\beta}}
v_{\underline{\beta}q}{}^\alpha =e^{\alpha q}\; , \qquad \hat{E}_{\beta\dot q}:= \hat{\cal E}^{\underline{\beta}}
v_{\underline{\beta}\beta\dot q} =
e^{\alpha q} h_{\alpha q\; \beta \dot q}+ e^b \chi_{b\; \beta\dot q}
\; . \qquad
\end{eqnarray}
With such a conventional constraints, the lowest dimensional (dim 0)
spin-tensorial component of the integrability condition for superembedding equation, Eq. (\ref{DEi=0}), reads $\gamma^i_{q\dot{q}}  h_{\beta p\; \alpha \dot q}+
\gamma^i_{p\dot{q}}  h_{\alpha q\; \beta \dot q}=0 $. The solution
of this equation is trivial, $h_{\alpha q\; \dot q}{}^{\!\beta}=0$,
so that Eqs. (\ref{Ef2=M20}) simplify to \cite{bpstv,HoweSezgin04}
\begin{eqnarray}\label{Ef2=M2}
\hat{E}^\alpha_q=e^{\alpha q}\; , \qquad \hat{E}_{\beta \dot q}=
 e^b \chi_{b\; \beta\dot q} \; . \qquad
\end{eqnarray}
Using  Eqs. (\ref{Ef2=M2}), the tangent superspace torsion
constraints (\ref{Ta=11D}), the conventional constraints resumed in
the first equation of (\ref{Dp:Du=Om}) and the superembedding
equation (\ref{Ei=0M2}), one finds that the bosonic torsion of the
worldvolume superspace reads
\begin{eqnarray}\label{Dea=M2}
De^a= 2i e^{\alpha q} \wedge e^{\beta q}\gamma^b_{\alpha\beta} - i
e^b\wedge e^c \, \chi_b\gamma^a \chi_c \; . \qquad
\end{eqnarray}

Now, the dimension 1/2, $\propto e^b \wedge e^{\alpha q}$ component
of Eq. (\ref{DEi=0}) express the spinorial component of Cartan form
$\Omega^{a i}\,$, $\; \Omega_{\alpha q}^{\;\; ai}= 2i
\gamma^i_{q\dot{p}} \chi^a_{\; \alpha \dot{p}} $; the dim 1,
$\propto e^b \wedge e^c$ component implies $\Omega_{[a\, b]}{}^i=0$,
which means that the pure bosonic component of the $\Omega^{a i}\,$
is symmetric, $\Omega_{b\, a}{}^i=\Omega_{(a\, b)}{}^i:=
u_{(a}{}^{\!\underline{\,c}}D_{b)}u_{\underline{c}}{}^i= -
D_{(a}u_{b)}{}^{\!\underline{\,c}}\; u_{\underline{c}}{}^i$ and coincides with  the
(superfield generalization of the) second fundamental form of the
worldvolume superspace considered as a surface in the target
superspace, Eq. (\ref{Kabi=M2}).

To resume, the dim 1/2 and 1 components of the integrability
conditions (\ref{DEi=0}) for the superembedding equation
(\ref{Ei=0M2}) gives us the expression for the generalized Cartan
form $\Omega^{ai}$ in terms of the second fundamental form
$K_{b}{}^{ai}$ of Eq. (\ref{Kabi=M2}), and in terms of the fermionic superfield
$\chi_{b\dot{q}}{}^{\beta} = \hat{E}_b^{\underline{\alpha}}\,
v_{\underline{\alpha}\dot{q}}{}^{\beta} :=
D_b\hat{Z}^{\underline{M}}\,
{\cal E}_{\underline{M}}{}^{\underline{\alpha}}(\hat{Z})\,
v_{\underline{\alpha}\dot{q}}{}^{\beta} $ which,  in
this sense, is a superpartner of the second fundamental form,
\begin{eqnarray}\label{Omai=M2}
&& \Omega^{ai}=   2i e^{\alpha q}
\gamma^i_{q\dot{p}}\chi^a{}_{\alpha \dot{p}} + e_b K^{ab\, i}\; ,
\qquad \\ \nonumber && \qquad K_{a\, b}{}^i:= -
 D_{(a}\hat{E}_{b)}{}^{\!\underline{\,c}}\; u_{\underline{c}}{}^i\; , \qquad
 \chi_{b\dot{q}}{}^{\beta} = \hat{E}_b^{\underline{\alpha}}\,
v_{\underline{\alpha}\dot{q}}{}^{\beta} := D_b\hat{Z}^{\underline{M}}\,
{\cal E}_{\underline{M}}{}^{\underline{\alpha}}(\hat{Z})\,
v_{\underline{\alpha}\dot{q}}{}^{\beta}\;  . \qquad
\end{eqnarray}

Now we turn to the selfconsistency conditions for the second
equation in (\ref{Ef2=M2}). It reads
\begin{eqnarray}\label{DEf=Dechi}
0=  D(\hat{E}_{\alpha\dot q}-
 e^b \chi_{b\, \alpha\dot q}) = \hat{T}^{\underline{\alpha}} v_{\underline{\alpha} \, \alpha
 \dot{q}} - {1\over 2} e^{\beta p} \wedge \Omega^{ai}\, \gamma_{a \, \alpha\beta}
 \gamma^i_{p\dot{q}} + i e^{\beta p} \wedge e^{\gamma p} \gamma^b_{\beta\gamma}\chi_{b\alpha
 \dot{q}} + \qquad \nonumber \\  + i e^b\wedge e^c \,
\chi_b\gamma^a \chi_c \, \chi_{a \beta\dot q} - e^b\wedge \, D
\chi_{b \beta\dot q}\; , \qquad
\end{eqnarray}
where we have used the expression for the bosonic torsion of the
worldvolume superspace (\ref{Dea=M2}), as well as the expression for
the derivative of the spinorial harmonic,
\begin{eqnarray}\label{Dvadq=M2}
 Dv_{\underline{\alpha} \, \alpha
 \dot{q}} = - {1\over 2} \Omega^{ai}\, v_{\underline{\alpha}}{}^{\beta
 p} \gamma^i_{p\dot{q}} \gamma_{a \, \alpha\beta}\; ,
\end{eqnarray}
which
appears as of the rectangular  blocks
of Eq. (\ref{DV=VOm}).

Taking into account the expression (\ref{Omai=M2}) for
$\Omega^{ai}$, one finds that the lowest dimensional, $\propto
e^{\beta p}\wedge e^{\gamma p^\prime}$ component of Eq.
(\ref{DEf=Dechi}) reads $ - i \gamma^i_{p\dot{q}}
\gamma^i_{p^\prime\dot{p}} \gamma^{a}_{\alpha \beta} \chi_{a \gamma
\dot{p}}- i\gamma^i_{p^\prime\dot{q}} \gamma^i_{p\dot{p}}
\gamma^{a}_{\alpha \gamma} \chi_{a \beta \dot{p}} + 2i\delta_{p\,
p^\prime} \chi_{a \, \alpha\dot{q}}=0$. The only consequence of this
equation is that $\gamma_{a \alpha \beta} \chi^a_{\gamma \dot{p}}-
\gamma_{a \alpha \gamma} \chi^a_{\beta \dot{p}} =0$ which is an
equivalent form of the fermionic equations
\begin{eqnarray}\label{fEqmM2}
\tilde{\gamma}^{a \alpha \beta} \chi_{a \beta \dot{q}} :=
\tilde{\gamma}^{a \alpha \beta} \hat{E}_a{}^{\underline{\alpha}}
v_{\underline{\alpha}\beta \dot{q}}=0 \;  . \qquad
\end{eqnarray}
Then, the dimension 1, $\propto e^{b}\wedge e^{\beta p}$ component
of Eq. (\ref{DEf=Dechi}) is $0=D_{\beta p} \chi_{b \alpha \dot{q}}+
v_{\beta p}{}^{\underline{\beta}}
\hat{T}_{\underline{\beta}\underline{a}}{}^{\underline{\gamma}}v_{\underline{\gamma}\;\alpha
\dot{q}}u_b{}^{\underline{a}} + {1\over 2}\gamma^i_{p\dot{q}}
\gamma^a_{\alpha\beta} K_{ab}{}^i$. Contracting this equation with
$\tilde{\gamma}^{b \gamma\alpha}$ one finds
\begin{eqnarray}\label{giKaai=}
\gamma^i_{p\dot{q}} K_{a}{}^{ai} \delta_\beta{}^\gamma = - 2
v_{\beta p}{}^{\underline{\alpha}}
\hat{T}_{\underline{\alpha}\underline{a}}{}^{\underline{\delta}}\,
v_{\underline{\delta}\;\alpha \dot{q}}\tilde{\gamma}^{b\alpha\gamma}
u_b{}^{\underline{a}} - 2 D_{\beta p}(\tilde{\gamma}^{a } \chi_{a
\dot{q}})^\gamma \; . \qquad
\end{eqnarray}
The last term vanishes due to the fermionic equation of motion
(\ref{fEqmM2}), so that
\begin{eqnarray}\label{giKaai=T}
\gamma^i_{p\dot{q}} K_{a}{}^{ai} \delta_\beta{}^\gamma = - 2
v_{\beta p}{}^{\underline{\alpha}}
\hat{T}_{\underline{\alpha}\underline{a}}{}^{\underline{\delta}}\,
v_{\underline{\delta}\;\alpha \dot{q}}\tilde{\gamma}^{b\alpha\gamma}
u_b{}^{\underline{a}}\; . \qquad
\end{eqnarray}
The bosonic equations of motion are obtained by contracting this
equation with $1/16 \gamma^i_{p\dot{q}}\delta_\gamma{}^\beta$. It
reads
\begin{eqnarray}\label{bEqmM2}
 K_{a}{}^{ai} := -D^a \hat{E}_a{}^{\underline{b}}\; u_{\underline{b}}{}^i = - {1\over 8} v_{\beta
p}{}^{\underline{\alpha}} \gamma^i_{p\dot{q}}
\tilde{\gamma}^{b\beta\alpha} v_{\underline{\delta}\;\alpha \dot{q}}
u_b{}^{\underline{a}}
\hat{T}_{\underline{\alpha}\underline{a}}{}^{\underline{\delta}}\,\;
. \qquad
\end{eqnarray}
The fact that other irreducible parts of the {\it r.h.s.} of Eq.
(\ref{giKaai=T}) vanishes, {\it i.e.} that  $v_{\beta
p}{}^{\underline{\alpha}}
\hat{T}_{\underline{\alpha}\underline{a}}{}^{\underline{\delta}}\,
v_{\underline{\delta}\;\alpha \dot{q}}\tilde{\gamma}^{b\alpha\gamma}
u_b{}^{\underline{a}} \propto \gamma^i_{p\dot{q}}
\delta_\beta{}^\gamma $, might contain a nontrivial information on the
geometry of the D=11 superspace supergravity background. One can
check that this is satisfied identically  for
\begin{eqnarray}\label{Tfbf}
{T}_{\underline{\beta}\underline{a}}{}^{\underline{\gamma}}= - {i
\over 144} \left(
F^{\underline{c_1}\underline{c_2}\underline{c_3}\underline{c_4}}
\Gamma_{\underline{a}\underline{c_1}\underline{c_2}\underline{c_3}\underline{c_4}}+
8F_{\underline{a}\underline{c_1}\underline{c_2}\underline{c_3}}\Gamma^{\underline{c_1}\underline{c_2}\underline{c_3}}\right){}_{\underline{\beta}}{}^{\underline{\gamma}}\;
,
 \qquad
\end{eqnarray}
which follows from the standard superspace constraints of D=11 supergravity \cite{CremmerFerrara80,BrinkHowe80} by studying the Bianchi identities.
Using (\ref{Tfbf}) one can obtain the more specific form of the (superfield) bosonic
equations of the M2-brane: Eq. (\ref{bEqmM2}) is equivalent to
\begin{eqnarray}\label{bEqM2=SEmb}
K_{a}{}^{ai} = {1\over 3}\; F^i{}_{abc}\varepsilon^{abc}\; , \qquad
\; ,     \qquad
\end{eqnarray}
where
\begin{eqnarray}\label{flux=iabc}
F^i{}_{abc}:=
F_{\underline{a}\underline{b}\underline{c}\underline{d}}(\hat{Z}) u^{\underline{a}i}
u_b{}^{\underline{b}}u_c{}^{\underline{c}}u_d{}^{\underline{d}}
 \; .    \qquad
\end{eqnarray}

\bigskip
To make a contact with standard formulation of the supermembrane
\cite{BST87}, let us notice that, on the bosonic worldvolume,
ignoring fermions,  and writing equations in terms of the induced
metric ($g_{mn}=e_m{}^ae_{an}=
\hat{E}_m{}^{\underline{a}}\hat{E}_{n\underline{a}}$), one finds that  $D^a \hat{E}_a{}^{\underline{b}}= D_m
(\sqrt{|g|}g^{mn}\hat{E}_n{}^{\underline{b}})\;$,  where $D_m$ is
the $SO(1,9)$ covariant derivative on the worldvolume. Hence,  Eq. (\ref{bEqM2=SEmb})
coincides in this case  with the standard supermembrane equation
\begin{eqnarray}\label{bEqM2=st}
D_m (\sqrt{|g|}g^{mn}\hat{E}_n{}^{\underline{b}})= - \, {1\over 3}\;
\eta^{\underline{b}\underline{a}}\, F_{\underline{a}\;
bcd}\varepsilon^{bcd}\; , \qquad F_{\underline{a}\;abc}:=
F_{\underline{a}\underline{b}\underline{c}\underline{d}}
\hat{E}_b{}^{\underline{b}}\hat{E}_c{}^{\underline{c}}\hat{E}_d{}^{\underline{d}}
 \; .    \qquad
\end{eqnarray}
contracted with the orthogonal harmonics $u_{\underline{a}}{}^i$ ($K_a{}^{ai}:= - D^a \hat{E}_a{}^{\underline{b}}
u_{\underline{b}}{}^i = D_m
(\sqrt{|g|}g^{mn}\hat{E}_n{}^{\underline{b}})u_{\underline{b}}{}^i\;$).
The projection of the supermembrane equation onto the vector
harmonics $u_{\underline{b}}{}^a$, tangential to the worldvolume, $D^a
\hat{E}_a{}^{\underline{b}} u_{\underline{b}}{}^b = ...$, can be shown to be satisfied
identically. This is the Noether identity reflecting the
reparametrization invariance of the supermembrane (action and of the) equations of motion.
Thus Eq. (\ref{bEqM2=SEmb}) is equivalent to the standard supermembrane equation, Eq. (\ref{bEqM2=st})
modulo fermionioc contributions.

Coming back, let us stress that Eq.  (\ref{giKaai=}) gives us the
interrelation between the fermionic and the bosonic equations, Eqs.
(\ref{fEqmM2}) and (\ref{bEqmM2}) of supermembrane in general D=11
superfield supergravity background. It shows that the bosonic equation of motion of the M2-brane can be obtained as a second component in the decomposition of the superfield generalization of the fermionic equation of motion on the Grassmann coordinate.

\subsection{M5-brane in superembedding approach}

The dynamics of M5-brane is also fixed by the superembedding equation (\ref{SembEq}) \cite{hs2} equivalent to (\ref{Ei=0}),
\begin{eqnarray}\label{Ei=0M5}
\hat{E}^{i}:= \hat{E}^{\underline{\, a}}u_{\underline{\, a}}{}^i = 0
\; .  \qquad
\end{eqnarray}
The bosonic supervielbein of the worldvolume superspace is defined by (\ref{Eua=ea}) and the worldvolume superspace $SO(1,5)$ and  $SO(5)$ connections - by (\ref{Dp:Du=Om}). The fermionic supervielbein of the M2-brane worldvolume superspace ${\cal W}^{(6|16)}$ can be
identified with, say, $\hat{E}^{\alpha q}:= {\hat{\cal E}}^{\underline{\beta}}
v_{\underline{\beta}}{}^{\alpha q}$.  Then,
\begin{eqnarray}\label{Ef1=M5}
\hat{E}^{\alpha q}&:= & \hat{\cal E}^{\underline{\beta}}
v_{\underline{\beta}}{}^{\alpha q} =e^{\alpha q}\; , \qquad \\ \label{Ef2=M5} \hat{E}_{\beta}^{q}\; &:=& \hat{\cal E}^{\underline{\alpha}}
v_{\underline{\alpha}}{}_{\beta}^{q} \; =
e^{\alpha q} h_{\alpha\beta} + e^b \chi_{b\; \beta}{}^{q}
\; . \qquad
\end{eqnarray}
To be more precise, the general decomposition of the second projection of the pull--back  of the target superspace fermionic supervielbein ${\cal E}^{\underline{\alpha}}$ reads $\hat{{\cal E}}^{\underline{\alpha}}
v_{\underline{\alpha}}{}_{\beta}{}^{\! q} =
e^{\alpha p} h_{\alpha p\; \beta}{}^q + e^b \chi_{b\; \beta}{}^{q}$. However, as the further study shows anyway that
$h_{\alpha p\; \beta}{}^q = h_{\alpha\beta}\delta_p{}^q$,  we have allowed ourself to make a shortcut substituting this expression in Eq.  (\ref{Ef2=M5}) from the very beginning.

Eqs. (\ref{Ef1=M5}) and (\ref{Ef2=M5})  can be collected in
\begin{eqnarray}\label{Ef=eV(h)+}
\hat{{\cal E}}^{\underline{\alpha}}=  e^{\beta q} V_{\beta q}{}^{\underline{\alpha}}(h) + e^a \chi_{a\beta}{}^p v_p{}^{\beta \underline{\alpha}}\; , \qquad  V_{\beta p}{}^{\underline{\alpha}}(h):=
 v_{\beta p}{}^{\underline{\alpha}}+  h_{ \beta\gamma } v_{p}{}^{\gamma \underline{\alpha}}\; .  \qquad
\end{eqnarray}
For the discussion below it is useful to notice that the `deformed harmonics' $V_{\beta p}{}^{\underline{\alpha}}(h):=
 v_{\beta p}{}^{\underline{\alpha}}+  h_{ \beta\gamma } v_{p}{}^{\gamma \underline{\alpha}}$ obey (see Appendix A2 for our notation $\Gamma$-matrices representation and $\gamma$--matrices properties)
\begin{eqnarray}\label{uuV(h)V(h)=}
u_a{}^{\underline{a}}u_b{}^{\underline{b}} V_{\beta p}{}^{\underline{\alpha}}(h) \Gamma_{\underline{a}\underline{b}\; \underline{\alpha}\underline{\delta}}V_{\beta p}{}^{\underline{\delta}}(h)= 2i (\gamma_{ab}h)_{[\alpha\beta]}C_{qp}\; .   \qquad
\end{eqnarray}

The lowest dimensional ($\propto e^{\alpha q}\wedge e^{\beta p}$) component of the integrability conditions for the superembedding equation, Eq. (\ref{DEi=0}), results in $h_{\alpha\beta}=h_{\beta \alpha}$. As in $d=6$ the basis of symmetric spin tensor matrix is provided by $\gamma^{abc}_{\alpha\beta} $ (notice that $\gamma^a_{\alpha\beta}=- \gamma^a_{\alpha\beta}= {1\over 2}\epsilon_{\alpha\beta\gamma\delta}\tilde{\gamma}^{a\gamma\delta}$ and $\gamma^{(a}\tilde{\gamma}{}^{b)}=\eta^{ab}$;  see \cite{PKT+6d} and Appendix A2 for more detail)  so that
\begin{eqnarray}\label{M5:h=h3g3}
   h_{\alpha\beta}= {1\over 3!} h_{abc}\gamma^{abc}_{\alpha\beta}\; .  \qquad
\end{eqnarray}
As far as $\gamma^{abc}_{\alpha\beta}$ is anti-self-dual, $\gamma^{abc}_{\alpha\beta}=- {1\over 3!}\epsilon^{abcdef}\gamma_{def \alpha\beta}$, the antisymmetric tensor $h_{abc}$ in (\ref{M5:h=h3g3}) is self-dual,
\begin{eqnarray}\label{M5:h=*h}
   h_{abc}= {1\over 3!} \epsilon_{abcdef} h^{def}\; .  \qquad
\end{eqnarray}
An important property of the symmetric spin-tensor $h_{\alpha\beta}$ is ({\it cf.} (\ref{hsah=ska}))
\begin{eqnarray}\label{M5:htgh=}
  h\tilde{\gamma}{}^ah= \gamma^bk_b{}^a\; , \qquad k_b{}^a= -2 h_{bcd}h^{cda}\;  .  \qquad
\end{eqnarray}
One easily obtains this taking into account that, as a consequence of the self--duality  (\ref{M5:h=*h}), the contraction of $h_{abc}$ with $\tilde{\gamma}{}_{abc}^{\alpha\beta}= + {1\over 3!} \epsilon_{abcdef} \tilde{\gamma}{}^{def\, \alpha\beta}$ vanishes. Then $h\tilde{\gamma}{}^ah=h^{abc}(\tilde{\gamma}_{bc}h)$ from which one easily arrives at (\ref{M5:htgh=}).

The appearance of a third rank antisymmetric self-dual tensor reflects the fact that the linearized spectrum of the M5-brane includes the chiral two-form potential \cite{Witten:1996hc} {\it i.e.} the two form 6d gauge field with the self-dual three form field strength. Beyond the linear approximation, one finds that the gauge field strength tensor obeys a nonlinear generalization of the self-duality condition \cite{hs2,blnpst,schw5}.

The dim 3/2 and dim 2 components of the integrability condition Eq. (\ref{DEi=0}) determines the generalized Cartan form to be
\begin{eqnarray}\label{M5:Omai=}
  \Omega^{ai}= 2e^{\alpha q}\gamma^i_{qp} \chi_\alpha{}^{a p} + e_b K^{ab\, i}  \; ,  \qquad
\end{eqnarray}
where $K^{ab\, i} =K^{ab\, i} $ is the second fundamental form defined as in Eq. (\ref{Kabi=M2}) and
$\gamma^i_{qp} =- \gamma^i_{qp} = {1\over 2}\epsilon_{qprs} \tilde{\gamma}^{i\, rs}=- (\tilde{\gamma}^{i\, qp})^* $ are the SO(5) Klebsh-Gordan coefficients (see Appendix A2 for their properties).

The bosonic torsion of the worldvolume geometry induced by superembedding reads
\begin{eqnarray}\label{M5:Dea=}
 De^a =- ie^{\alpha q}\wedge e^{\beta p} C_{qp} \gamma^b_{\alpha\beta}\, m_b{}^a\; + 2i e^{b}\wedge e^{\alpha q} C_{qp} (h\tilde{\gamma}{}^a\chi_{b}{}^p)_\alpha + i e^c\wedge e^b \psi_b^q \tilde{\gamma}{}^a\psi_c^p C_{qp} \; ,  \qquad
\end{eqnarray}
where \cite{hs2,hsw}
\begin{eqnarray}\label{M5:mab=}
m_a{}^b=\delta_a{}^b+ k_a{}^b= \delta_a{}^b- 2h_{acd}h^{bcd}
\; .   \qquad
\end{eqnarray}
Generically, this matrix  is invertible (and not $k$ of (\ref{M5:htgh=}); {\it cf.}   Eq. (\ref{hsah=ska}) in the case of D$p$-branes).

Now we could pass to studying the selfconsistency condition for the fermionic one-form equation (\ref{Ef2=M5}),
\begin{eqnarray}\label{M5:0=DE2-}
& 0=D(\hat{E}_{\beta}{}^{\! q}-
e^{\alpha q} h_{\alpha\beta} - e^b \chi_{b\; \beta}{}^{\! q})= \hat{T}{}^{\underline{\alpha}} v_{\underline{\alpha}\beta}{}^{\! q}-{i\over 2} e^{\alpha p}\wedge \Omega^{ai} \gamma_{a\alpha\beta}(\gamma^iC)_p{}^q  - e^{\alpha q} \wedge D h_{\alpha\beta} - \nonumber \\ & - e^b
\wedge  D\chi_{b\; \beta}{}^{\! q} - De^{\alpha q} \, h_{\alpha\beta} - De^b \,\chi_{b\; \beta}{}^{\! q} \;  ,  \qquad\end{eqnarray}
and obtain all the dynamical equation from this. In the second equality of (\ref{M5:0=DE2-})  we have used the second  of the following two spinorial counterpart of Eqs. (\ref{Dp:Du=Om}),
\begin{eqnarray}\label{Dv=M5}
Dv_{\underline{\alpha}}{}^{\alpha q}= {i\over 2}\Omega^{ai}  v_{\underline{\alpha} \beta}{}^p
\tilde{\gamma}_a^{\beta\alpha} ({\gamma}^iC)_p{}^q \; , \qquad   Dv_{\underline{\alpha}\alpha }{}^{q}= - {i\over 2}\Omega^{ai}  v_{\underline{\alpha}}^{\beta p}
\tilde{\gamma}_{a\, \beta\alpha} ({\gamma}^iC)_p{}^q \; , \qquad
 \qquad
\end{eqnarray}
while the first one has to be used in calculation of fermionic torsion. Clearly, neither this nor the equation (\ref{M5:0=DE2-}) as a whole looks simple in general type II supergravity background.

However, the study may be simplified essentially if we use the presence of the above mentioned  two--form gauge field on the M5 worldvolume, generalize it to the superform $b_2$ on the worldvolume superspace, impose the constraints on its generalized field strength and study the corresponding Bianchi identities. This is the counterpart of imposing the gauge field constraints on the worldvolume superspace of D$p$-branes  which we discussed in Sec. 3.

The constraints on the 3-form field strength \cite{hs2}  can be written in the form
\begin{eqnarray}\label{M5:H3=db-=}
H_3:=db_2 - \hat{C}_3 = {1\over 3!}e^c\wedge e^b\wedge e^a H_{abc}
\; ,   \qquad\end{eqnarray}
where $\hat{C}_3$ is the pull--back to ${\cal W}^{(6|16)}$ of the three form gauge potential of the superspace 11D supergravity the field strength of which obeys the constraints
\begin{eqnarray}
\label{cF4=}
{\cal F}_4 &:=& dC_3={1\over 4} E^{\underline{b}} \wedge E^{\underline{a}} \wedge {\cal E} \wedge {\Gamma}_{\underline{a}\underline{b}}{\cal E}
 + {1 \over 4! } E^{\underline{d}}
\wedge \ldots \wedge E^{\underline{a}}  F_{\underline{a}\underline{b}\underline{c}\underline{d}} \; . \qquad
\end{eqnarray}
The Bianchi identities
\begin{eqnarray}
\label{M5:dH3=-F4}
dH_3=-{\cal F}_4 \;
\qquad
\end{eqnarray}
result in the relation between the tensor field strength $H_{abc}$ and self-dual tensor $h_{abc}$ of Eqs.  (\ref{M5:h=h3g3}), (\ref{M5:h=*h}) \cite{hs2,hsw}
\begin{eqnarray}
\label{mH=h}
m_a{}^dH_{bcd}=h_{abc}= {1\over 3!}\epsilon_{abcdef}h^{def}\;
\qquad
\end{eqnarray}
as well as
\begin{eqnarray}
\label{M5:DfH=}
D_{\alpha q}H_{abc} &=& - 6i C_{qp}(h\tilde{\gamma}^d\psi_{[a}) H_{bc]d}\; ,\qquad \\ \label{M5:DbH=}
D_{[a}H_{bcd]} &=& -3i C_{qp}(\psi_{[a}\tilde{\gamma}^e\psi_{b})H_{cd]e} +{1\over 4}u_a{}^{\underline{a}} \ldots u_d{}^{\underline{d}}F_{\underline{a}\underline{b}\underline{c}\underline{d}}(\hat{Z})\; .
\qquad
\end{eqnarray}
Clearly, Eqs. (\ref{mH=h}), in the derivation of which one uses the identity (\ref{uuV(h)V(h)=}),  provides a nonlinear generalization of the selfduality equation and, hence, imply dynamical equations of motion for the two--form  gauge field $b_2$. (To convince that this is the case, it is sufficient to note that the standard self-duality implies that the linearized 2-form gauge field equations of motion in d=6 are satisfied).

The above relatively simple derivation of the nonlinear selfduality equation (\ref{mH=h}) gives one more example of the usefulness of introducing the worldvolume superspace gauge potentials  and studying the corresponding Bianchi identities for their constrained field strengths. The details on derivation of the dynamical equations for the M5-brane coordinate functions from the superembedding description  can be found in the original articles \cite{hs2,hsw,blnpst2} and in the review \cite{Dima99}.
The proof of their equivalence to the equations of motion derived from the worldvolume action \cite{blnpst,schw5} is the subject of \cite{hsw,blnpst2}.

\section{Multiple D0-brane equations from superembedding approach.}
\label{multipleD0}

It is the usual expectation that the action for a system of N
D$p$-branes  will essentially be  a nonlinear generalization of the
U(N) SYM action. In particular, the (purely bosonic and not Lorentz
invariant) Myers action \cite{Myers:1999ps} is of this type. Then
the equations of motion which should follow from a hypothetical
supersymmetric and Lorentz covariant generalization (or
modification) of this action are expected to contain the SU(N) SYM
equations ($U(N)=SU(N)\times U(1)$) while the center of mass motion
is expected to  be described by a usual type of coordinate functions
$\hat{Z}^{\underline{M}}(\xi)$ and by related equations for the U(1)
gauge fields (presumably coupled to the SU(N) equations). Notice
that the center of mass equations of motion (and equations for U(1)
gauge fields which is expected to be involved in the center of mass
supermultiplet) are expected to be quite close to the equations  for
a single D$p$-brane, but with the single brane tension (mass) $T$
replaced by $NT$. In this section we review, following \cite{IB09:D0}, the application of the
superembedding approach in search for such a supersymmetric equations.

 \subsection{Non-Abelian
${\cal N}=16$, $d=1$ SYM constraints on D0-brane}

In \cite{IB09:D0} the worldvolume superspace of multiple D$0$-brane system was assumed to obey the same superembedding equation (\ref{Ei=0}) as in the case of single D--brane.

To motivate this, let us notice that the superembedding equation is pure geometrical. It is stating, in its form of
(\ref{SembEq}), that the pull--back of the target space bosonic
vielbein to the worldvolume superspace ${\cal W}^{(1|16)}$ do not
have projections on the fermionic vielbein of ${\cal W}^{(1|16)}$. Hence
it is natural to assume that the center of mass motion of the system
of multiple D0-brane will also obey the superembedding equations.

Of course this is not a proof. But the universality of the
superembedding equation, which is valid for all extended objects
studied till now in their maximal worldvolume superspace
formulations, and the difficulties one arrives at in any attempt to
modify to try to impose it, following \cite{IB09:D0}, at least as an approximation (see concluding Sec. 6 for more discussion on this).

As far as the superembedding equation puts the $p<6$ D$p$--brane models on
the mass shell, our superembedding approach to $p<6$  ND$p$-brane model
predicts that the center of mass motion will be described the motion
of single brane with tension  $N\cdot T$.   Then, in the light of
the above stated, and taking in mind that a good low energy approximation to mutiple D$p$-brane is given by maximally supersymmetric $d=p+1$ U(N) SYM action,  the only possibility to describe the multiple
D0-brane system in the framework of superembedding approach seems to
be to consider a {\it non-Abelian SU(N) gauge field supermultiplet}
on the D0-brane worldvolume superspace $W^{(1|16)}$. (See
\cite{IB08:Q7} for more discussion on a similar issue in the context of searching for
hypothetical Q7-branes \cite{Dima+Eric=2007}.)

 This can be defined by an $su(n)$ valued non-Abelian gauge potential
one form $A=e^0A_0 + e^\alpha A_\alpha$ with the field strength
\begin{eqnarray}\label{D0:G=dA-=}
G_2= dA - A\wedge A= {1\over 2}e^\alpha\wedge e^\beta
G_{\alpha\beta} + e^0\wedge e^\beta G_{\beta 0}\;  \qquad
\end{eqnarray}
which obeys the Bianchi identities
\begin{eqnarray}
 \label{ND0:BI=} DG_2= dG_2 - G_2\wedge A + A \wedge G_2\equiv 0 \; . \qquad
\end{eqnarray}
As in the Abelian case discussed in sec. 3, to get a nontrivial consequences for the structure of the field
strengths $G_{\alpha\beta}$, $G_{\beta 0}$ form Bianchi identities one has to impose
constraints. A natural possibilit
y  is
\begin{eqnarray}\label{D0:G=sX}
G_{\alpha\beta}= i \sigma^i_{\alpha\beta} {\bb X}^i \; , \qquad
\end{eqnarray}
with some $su(N)$ valued $SO(9)$ vector superfield ${\bb X}^i$. (See sec. 5.5 for discussion on possible modification of this constraint). The
Bianchi identities (\ref{ND0:BI=})
 are satisfied if  ${\bb X}^i$ obeys
\begin{eqnarray}\label{D0:DX=sPsi}
D_{\alpha} {\bb X}^i = 4i (\sigma^0\tilde{\sigma}{}^i)_{\alpha}{}^{
\beta} \, {\Psi}_\beta \; . \qquad
\end{eqnarray}
and $
G_{\alpha 0}= i\Psi_\alpha + {i\over 2} (\sigma^{0i}\Lambda)_\alpha
{\bb X}^i$.
It is natural to call (\ref{D0:DX=sPsi}) {\it superembedding--like
equation} as it gives a matrix $SU(N)$ gauge invariant generalization of the
gauge fixed form of the linearized superembedding equation
(\ref{SembEq}) (this reads  $D_\alpha X^i= \propto
(\sigma^0\tilde{\sigma}^i(\Theta^2-\Theta^1))_\alpha$, see
\cite{hs96}).

\subsection{Multiple D0-brane equations of motion from $d=1$ ${\cal N}=16$ SYM constraints. Flat target superspace. }
Let us, for simplicity, consider the case of flat target type IIA
superspace, in which, on the mass shell of D0-brane, $\Omega^i=0$,
so that $\sigma^0_{\alpha\beta}$ and $\sigma^i_{\alpha\beta}$ are
covariantly constants, $D\sigma^0_{\alpha\beta}=0
=D\sigma^i_{\alpha\beta}$. In this case the integrability conditions
($D_{(\beta}D_{\alpha )}{\bb X}^i=...$) for Eq. (\ref{D0:DX=sPsi})
result in
\begin{eqnarray}\label{D0:DPsi=}
D_{\alpha}\Psi_{\beta } = & -{1\over 2} \sigma^i_{\alpha\beta}
D_0{\bb X}^i+{1\over 16} \sigma^{0ij}_{\alpha\beta}
 [{\bb X}^i\, , \, {\bb X}^j ]\;   \qquad
\end{eqnarray}
and the integrability conditions for Eq. (\ref{D0:DPsi=}), result
in 1d Dirac equation of the form\footnote{An important check on
consistency is that the irreducible $\propto
\sigma_{\underline{a}_1\ldots \underline{a}_5}$ part of this
integrability conditions is satisfied identically; its $\propto
\sigma^0$ part gives (\ref{D0:suDirac}), while $\propto \sigma^i$
part gives (\ref{D0:suDirac}) times $\sigma^{0i}$. }
\begin{eqnarray}\label{D0:suDirac}
& D_{0}\Psi_{\beta}+ {1\over 4} [ (\sigma^{0j}\Psi)_{\beta}\, , \,
{\bb X}^j ]=0 \; .  \qquad
\end{eqnarray}
Applying the Grassmann covariant derivative $D_\alpha$ to the
fermionic equations (\ref{D0:suDirac}), one derives, after some
algebra, the following set of equations
\begin{eqnarray}\label{D0:D0D0Xi=}
D_0D_0{\bb X}^i - {1\over 32} [[{\bb X}^i\, , \, {\bb X}^j ]\, , \,
{\bb X}^j ]+ {i\over 8} \{ \Psi_\alpha , \Psi_\beta
\} \,\tilde{\sigma}^{i \alpha\beta}=0 \; , \qquad \\
\label{D0:D0XiXi=} [D_0{\bb X}^i , {\bb X}^i ] - 4i \{ \Psi_\alpha ,
\Psi_\beta \} \,\tilde{\sigma}^{0\alpha\beta}=0 \; .  \qquad
\end{eqnarray}
Eq. (\ref{D0:D0D0Xi=}) is a candidate bosonic equation of motion of multiple D0-brane system.
Eq.  (\ref{D0:D0XiXi=}) has the meaning of Gauss low which appears in gauge theories as an equation of motion for the time component of gauge potential (which usually plays the
r\^ole of Lagrange multiplier).

\subsection{Relation to D=10 SYM and M(atrix) model}

The appearance of the counterpart of Gauss low (\ref{D0:D0XiXi=}),
characteristic of gauge theory,   is not occasional. The point is
that our equations appear to be the $D=10$ SYM equations
dimensionally reduced to $d=1$. The reason is that  our constraints
(\ref{D0:G=sX}) for $d=1$, ${\cal N}=16$ SYM multiplet can be
obtained as a result of dimensional reduction of $D=10$
supersymmetric gauge theory. Indeed, the standard $D=10$ SYM constraints
imply vanishing of spinor-spinor component of the field strength,
\begin{eqnarray}\label{SYM10D=c}{\bb F}_{\alpha\beta}:= 2{\bb D}_{(\alpha} {\bb A}_{\beta )} + \{
{\bb A}_{\alpha},  {\bb A}_{\beta }\}- 2i
\sigma^{\underline{a}}_{\alpha\beta}{\bb A}_{\underline{a}}=0\; .  \qquad
\end{eqnarray}
Assuming independence of fields on the nine spacial coordinate, one
finds that spacial components ${\bb A}_{i}$ of the ten-dimensional
field strength are covariant and can be treated as scalar fields
\begin{eqnarray}\label{Ai=Xi/2} {\bb A}_{i}={\bb X}^i/2\;  . \qquad
\end{eqnarray}  Then the minimal covariant field
strength for $d=1$ SYM can be defined as
$ G_{\alpha\beta}:= 2{\bb
D}_{(\alpha} {\bb A}_{\beta )} +  \{ {\bb A}_{\alpha},  {\bb
A}_{\beta }\} - 2i \sigma^{\underline{0}}_{\alpha\beta}{\bb
A}_{\underline{0}}$  and, due to the original D=10 SYM constraints (\ref{SYM10D=c}),
this is equal to $i\sigma^i{\bb X}^i$, as in Eq. (\ref{D0:G=sX}),
\begin{eqnarray}\label{Gff:=} G_{\alpha\beta}:= 2{\bb
D}_{(\alpha} {\bb A}_{\beta )} +  \{ {\bb A}_{\alpha},  {\bb
A}_{\beta }\} - 2i \sigma^{\underline{0}}_{\alpha\beta}{\bb
A}_{\underline{0}}=  i \sigma^i_{\alpha\beta} {\bb X}^i \; .   \qquad
\end{eqnarray}

The above observation is important, in particular, because it indicates the
relation with Matrix model \cite{Banks:1996vh}. Indeed, this is  described by the Lagrangian obtained by  dimensional reduction of the $D=10$ SYM down to $d=1$ \cite{Banks:1996vh}. Actually, the $d=1$ dimensional reduction of the $U(N)$ $D=10$ SYM was the first model used to describe D0-brane
dynamics in \cite{D0-1996} even before the nonlinear DBI+WZ  action for
super-Dp-branes where constructed in \cite{Dpac,Dpac2,B+T=Dpac}. Our superembedding approach description \cite{IB09:D0} differs from the above mentioned $U(N)$ SYM approximation by that it uses the $SU(N)$ SYM to describe the  relative motion of the constituent branes,  while the $U(1)$ gauge field entering the multiplet describing the motion of the center of mass obeys the nonlinear Born-Infeld type equations; also the coordinate function describing the embedding of worldline superspace into the target superspace obey the nonlinear equations.  Even if such a manifestly supersymmetric and Lorentz covariant description appeared to be only approximate,  this would be a wider applicable approximation the use of which might be productive.

In the light of identification (\ref{Ai=Xi/2}) it becomes clear that the superembedding--like equation for the $SU(N)$--valued superfield ${\bb X}{}^i$ (\ref{D0:DX=sPsi}) comes from the consequence ${\bb F}_{\alpha \underline{a}}=2i (\sigma_{\underline{a}} \tilde{\Psi} )_{\alpha}$ (with ${\Psi}$ defined by $\tilde{\Psi}=:\tilde{\sigma}^0{\Psi}$) of the constraints (\ref{SYM10D=c}) and thus provide the {\it general} solution of the Bianchi identity
\begin{eqnarray}\label{D0:DG=0f}
I_{\alpha\beta\gamma }:=D_{(\alpha} G_{\beta\gamma)} +
t_{(\alpha\beta}{}^\delta G_{\gamma)\delta} +
4i\sigma^0_{(\alpha\beta}G_{\gamma)0}=0  \;  \qquad
\end{eqnarray}
in the present of these constraints.

To resume, for the multiple D0-brane system in flat target type IIA superspace the worldline
superspace ${\cal W}^{(1|16)}$ is flat and our superembedding approach results in equations which are equivalent to the ones obtained as a result of dimensional reduction of D=10 SYM. However, it can also be used to
describe the multiple D0-brane system  in curved supergravity
background, where the way through 10D SYM dimensional reduction is
obscure.

\subsection{Multiple D0-branes in curved type IIA background. Polarization by external fluxes. }

In the case of worldvolume superspace of D0--brane moving in curved
target type IIA superspace the calculations become more complex due to the presence of bosonic and fermionic background superfields. For instance, instead of (\ref{D0:DPsi=}),  one finds
\begin{eqnarray}\label{D0:DPsi=IIA}
& D_{\alpha}\Psi_{\beta}  =  -{1\over 2} \sigma^i_{\alpha\beta} + {1\over 16}
\sigma^{0ij}_{\alpha\beta}[{\bb X}^i\, , \, {\bb X}^j ] +
\hat{\Lambda}_{1\epsilon}\Psi_\delta \Sigma_{1}{}^{\epsilon\delta}{}_{\alpha\beta}
+ (\hat{\Lambda}_{2}\sigma^0)_{\epsilon}\Psi_\delta \Sigma_{2}{}^{\epsilon\delta}{}_{\alpha\beta}
\;   \qquad
\end{eqnarray}
with spin-tensors $\Sigma_{1,2}{}^{\epsilon\delta}{}_{\alpha\beta}$ possessing the properties $\sigma^{\underline{a}\underline{b}}{}_{\delta}{}^{\alpha} \Sigma_{1,2}{}^{\epsilon\delta}{}_{\alpha\beta}\propto \sigma^{\underline{a}\underline{b}}{}_{\beta}{}^{\epsilon}$ and
$D_\gamma \Sigma_{1,2}{}^{\epsilon\delta}{}_{\alpha\beta} \propto \Lambda$.
We will not need an explicit form of these (we leave this and other details for future publication) as our main interest here will be in the algebraic structure of the bosonic  equations of motion (see Appendix C for the structure of fermionic equations).
Up to the fermionic bilinears  proportional to the fermionic background fields these bosonic equations read
\begin{eqnarray}\label{D0:D0D0Xi=IIA}
 D_0D_0{\bb X}^i - {1\over 32} [[{\bb X}^i\, , \, {\bb X}^j ]\, ,
\, {\bb X}^j ]+ {i\over 8} \{ \Psi_\alpha , \Psi_\beta \}
\,\tilde{\sigma}^{i \alpha\beta}=  D_0{\bb X}^j \, {\bb F}^{j,i} +
{1\over 16}[{\bb X}^j\, , \, {\bb X}^k]\,   {\bb G}^{jk,i} + \qquad \\
\nonumber + {\cal O}(\hat{\Lambda}_{1,2}\cdot \Psi)+ {\cal
O}(\hat{\Lambda}_{1,2}\cdot \hat{\Lambda}_{1,2})\; . \qquad
\end{eqnarray}
The SO(9) tensor coefficients   ${\bb F}^{j,i}$ and $ {\bb G}^{jk,i} $ in the {\it r.h.s.} of
(\ref{D0:D0D0Xi=IIA}) are expressed in terms of the NS-NS and RR fluxes by
\begin{eqnarray}\label{D0:FFij=}
{\bb F}^{j,i} = q_0 \widehat{D_0\Phi}\delta^{ij} + p_{1}
\hat{R}^{ij}+ q_{2} \hat{H}{}^{0ij}\; . \qquad
\\ \label{D0:GGijk=}  {\bb G}^{jk,i}= p_0\delta^{i[j}\widehat{D^{k]}\Phi}+ q_1 \delta^{i[j}\hat{R}{}^{k]0}+ p_2 \hat{H}{}^{ijk}+
q_3 \hat{R}{}^{0ijk}\; , \qquad
\end{eqnarray}
Here $q_{0,1,2,3}$ and $p_{0,1,2}$ are constant coefficients
characterizing couplings to dilaton as well as to electric and
magnetic fields strength of the 1-form, 2-form and 3-form gauge fields;
the RR field strength are defined by $R_{2n+2}=dC_{2n+1}- C_{2n-1}\wedge H_3$ and the 3-form field strength  of the NS-NS 2-form gauge field is simply $H_3=dB_2$.

Notice that the center of mass motion is factored out and is
described by the single D0-brane equations (\ref{D0:bEqm=}),
\begin{eqnarray}\label{D0:bEqmDDXi=}
K^i:= D_0D_0\hat{X}^i + ... =
e^{\hat{\Phi}}\hat{R}^{0i}+\widehat{D^i\Phi}+ {\cal O}(fermi^2)\; , \qquad
\end{eqnarray}
where $\hat{X}^i:=
\hat{Z}^ME_M^{\underline{a}}(\hat{Z})u_{\underline{a}}{}^i=\hat{X}^{\underline{a}}
u_{\underline{a}}{}^i+ ... $. Comparing Eq. (\ref{D0:bEqmDDXi=}) with Eq.
(\ref{D0:D0D0Xi=IIA}) we see that the multiple D0-branes, as
described by this equation, acquire interaction with higher form
'electric' and 'magnetic' fields
$\hat{H}^{0ij}:=H_{\underline{a}\underline{b}\underline{c}}(\hat{Z})u^{\underline{a}0}u^{\underline{b}i}u^{\underline{c}j}$,
$H^{ijk}:=H_{\underline{a}\underline{b}\underline{c}}(\hat{Z})u^{\underline{a}i}u^{\underline{b}j}u^{\underline{c}k}$,
$\hat{R}^{0ijk}:=R_{\underline{a}\underline{b}\underline{c}\underline{d}}(\hat{Z})
u^{\underline{a}0}u^{\underline{b}i}u^{\underline{c}j}u^{\underline{d}k}$.
As one D$0$-brane does not interact with these background, one may
say that the multiple D0-brane system is 'polarized' by the external
fluxes such that the interaction with higher brane gauge fields is
induced, much in the same way as neutral dielectric is polarized
and, due to this polarization, interacts with electric field. This
is the famous 'dielectric brane' effect first observed by Emparan \cite{Emparan:1997rt} and then by Myers in his purely bosonic nonlinear action \cite{Myers:1999ps}.

\subsection{Possible deformation of the constraints and superembedding equations}

The relation of our description of multiple D$0$-brane system with the dimensional reduction of $SU(N)$  SYM model suggests a possible existence of modifications of our $d=1$ ${\cal N}=16$ SYM constraints (\ref{D0:G=sX}). What one can certainly state is that such a modification exists for the case of multiple D$0$-brane system in flat target superspace.

Indeed, according to \cite{Cederwall:2001td,Movshev:2009ba} the most general deformation of the $D=10$ SYM constraints by contributions of the fields of SYM supermultiplet at the order $(\alpha^\prime)^2$ reads\footnote{The author thanks Linus Wulff for useful discussions on the SYM deformations.}
\begin{eqnarray}\label{SYM10D=c}{\bb F}_{\alpha\beta}= {\beta} (\sigma^{\underline{a}}\tilde{\Psi})_\alpha (\sigma^{\underline{b}}\tilde{\Psi})_\beta {\bb F}_{\underline{a}\underline{b}}\; ,  \qquad
\end{eqnarray}
where $\beta$ is a constant proportional to the second power of the Regge slop parameter, $\beta\propto (\alpha^\prime)^2$, and $\tilde{\Psi}$ is the basic superfield strength of the $D=10$ SYM multiplet. This appeared in
the equation ${\bb F}_{\alpha\underline{a}}= 2i(\sigma_{\underline{a}}\tilde{\Psi})_\alpha$ which follows from the standard SYM constraints ${\bb F}_{\alpha\beta}=0$. Of course, when the dim 1 constraint becomes (\ref{SYM10D=c}), the dim 3/2 equation also gets modified by $\propto \beta$ contributions, ${\bb F}_{\alpha\underline{a}}= 2i(\sigma_{\underline{a}}\tilde{\Psi})_\alpha + {\cal O}(\beta )$.

The dimensional reduction of the deformed SYM theory characterized by the constraints (\ref{SYM10D=c}) implies the following constraints  for the minimal field strength of the dimensional reduced $d=1$ theory
\begin{eqnarray}\label{Gff:=} G_{\alpha\beta}=  i\sigma^i_{\alpha\beta}{\bb X}^i - \beta {\Psi}_{(\alpha} (\sigma^{0i}{\Psi})_{\beta )}D_0{\bb X}^i +  {\beta\over 4} (\sigma^{0i}{\Psi})_{\alpha} (\sigma^{0j}{\Psi})_\beta
[{\bb X}^i , {\bb X}^j]\;   , \qquad \\
\label{P=stP} {\Psi}={\sigma}^0\tilde{\Psi}\; . \qquad
\end{eqnarray}
This can be used now as a constraint for $d=1$, ${\cal N}=16$  SYM model leaving on the worldline of a D$0$-brane moving in flat targets superspace (as such a superspace is flat). This, in its turn, implies the following modification of the superembedding--like equation (which can be obtained by dimensional reduction of the consequence ${\bb F}_{\alpha\underline{a}}= 2i(\sigma_{\underline{a}}\tilde{\Psi})_\alpha + {\cal O}(\beta )$ of the modified constraint (\ref{SYM10D=c})),
\begin{eqnarray}\label{Gff:=} D_{\alpha} {\bb X}^i = 4i (\sigma^{0i}\Psi)_\alpha
+ 3i \beta \tilde{\sigma}{}^{i\, \beta \gamma} D_{(\alpha}\left({\Psi}_{\beta} (\sigma^{0i}{\Psi})_{\gamma )}D_0{\bb X}^i +  {1\over 4} (\sigma^{0i}{\Psi})_{\beta} (\sigma^{0j}{\Psi})_{\gamma )}
[{\bb X}^i , {\bb X}^j]\right)\; . \qquad
\end{eqnarray}

Even leaving aside the question of whether a counterpart of such modification can be found for the case of D$0$--brane worldvolume moving in an arbitrary curved type IIA supergravity background, one sees that these constraints are too complex. It is very hard to deal with them, at least without use of a computer programm (see \cite{Cederwall:2001dx} for an efficient use of computer programmes in superfield calculations).

Then, even if our formulation of superembedding approach to multiple D-brane system based on superembedding and superembedding--like equation as well as on the constraints (\ref{D0:G=sX}) is approximate, it promises to be an efficient approximation to study such a systems. Following \cite{IB09:D0}, we have proved that  such approach  exists and  is consistent in the case of multiple D-particle (D$0$-brane) system. An important problem is to understand whether it can be extended to type IIB multiple D-strings (D$1$-branes), D-membrane (D$2$--brane) and higher D$p$-brane systems.

\section{Conclusion and discussions}
\setcounter{equation}{0}

In this contribution we review superembedding approach to D-branes and M-branes
\cite{bpstv,hs96,hs2}
as well as its recent application \cite{IB09:D0} to searching for the covariant and supersymmetric equation for multiple D-brane systems.

We begin by general review of the superembedding approach to D$p$-brane, which happens to be simpler because, at least on the level of details of present contribution, it does not require introducing the spinor moving frame variables (see \cite{IB+DS06} where one can see the stage on which the introduction of these variable  is hardly possible without breaking the Lorentz invariance). Then we review superembedding approach to M2- and M5-brane, where the spinor moving frame variables do play essential r\^ole. In our review of superembedding description of  D- and M-branes we put an emphasis first on the universality of the superembedding equation which, for the most interesting cases of M2-, M5- and D$p$-branes with $p<6$ specify completely not only the worldvolume superspace geometry but also the dynamics of the brane. We also stressed the usefulness of introducing the worldvolume superspace gauge forms corresponding to the worldvolume gauge fields and studying the Bianchi identities for their constrained field strength. This is inevitable for D$p$-branes with $p>5$ but also very convenient for the branes the dynamics of which is completely specified by superembedding equation. The superfield description of the worldvolume gauge fields for a single D-brane (and chiral two form gauge field of M5-brane) suggests to try to describe a multiple D$p$--brane system by putting an additional non-Abelian $SU(N)$ gauge supermultiplet, described by a set of worldvolume superspace constraints, on the worldvolume superspace of a single D$p$-brane.

In sec. 5  we, following \cite{IB09:D0}, apply
superembedding approach to search for the multiple D$0$--brane
equations on this line. We show that for the case of arbitrary (on-shell) type II supergravity
background the dynamical equations obtained from the superembedding
approach describe the coupling of multiple D$0$--branes to the
higher NS-NS and RR fluxes ($H^{0ij}$, $H^{ijk}$ and $R^{0ijk}$).
Thus our equations of motion show the 'polarization' of multiple
D0-brane system which generates charges characteristic for higher
D-brane. This is the content of the so-called 'dielectric brane
effect' \cite{Emparan:1997rt,Myers:1999ps} characteristic for the (purely bosonic)
Myers action \cite{Myers:1999ps}. Further study of these equations
and of possible restrictions which they might put on the pull--back
of background fluxes to the worldline is an interesting problem for
future study.

In the case of flat tangent superspace, when the background fluxes vanish, the d=1, {\cal N}=16 worldvolume superspace of D0--brane is flat and the  dynamical equations for the relative motion of D$0$-brane 'constituents', which follows from the superembedding approach,  are those of the D=10 SU(N) SYM dimensionally reduced down to $d=1$.  They, thus, essentially coincide with what had been
used for the  very low energy description of multiple D0--brane system \cite{D0-1996} and with the Matrix model \cite{Banks:1996vh}.

The purely bosonic limit of our equations is clearly simpler than the equations following from the Myers action \cite{Myers:1999ps}. It is tempting to propose that this simpler but covariant and manifestly supersymmetric equations, together with the single D0--brane equation describing the center of mass motion, actually give the 'complete' description of the multiple D0-brane system \cite{IB09:D0}.
Furthermore, as we have already stressed, these give the completely supersymmetric and Lorentz invariant description of the 'dielectric brane effect'. The advantage of this description is that it is supersymmetric and also Lorentz invariant, while the Myers proposal \cite{Myers:1999ps} does possess neither of these symmetries expected for a system of multiple  Dp-branes\footnote{Let us recall that the Myers action was (and is) motivated by that it is derived from
T-duality. The starting point for the
corresponding chain of duality transformations is the
purely--bosonic D=10 non-Abelian Born-Infeld action based on the
symmetric trace prescription \cite{Tseytlin:BI-DBI} for the ordering of the SYM field
strength operators. Notice, however, that  supersymmetric
generalization of these 10D symmetric trace BI action is not known.}.

However, the existence of the deformation of our equations for the case of multiple D$0$ in flat target type IIB superspace, which follows from the existence of the deformation of the 10D SYM equations in flat $D=10\; {\cal N}=1$ superspace, suggests to allow the possible existence of deformation of our equations. However, one sees that the deformed  multiple D$0$ equations in flat target type IIB superspace, the explicit form of which is presently available,  are very complicated and its use looks inefficient (at least without the using computer programs).

Then, even if approximate, our superembedding description based on superembedding and superembedding-like equation plus simplest gauge field constraints, might provide useful approximation of nearly-coincident multiple branes, which goes beyond the $U(N)$ SYM description as far as the fields related to the center of mass motion are allowed to be strong.

As we have mentioned in the text, a very interesting  boundary fermion approach to the description of multiple D$p$-branes was developed by Howe, Lindsrom and Wulff in \cite{Howe+Linstrom+Linus,Howe+Linstrom+Linus=2007}. Presently the top-line result of this approach is the supersymmetric action possessing the kappa symmetry on the classical (or 'minus one quantization') level, {\it i.e.} before quantizing boundary fermions \cite{Howe+Linstrom+Linus=2007}. However, the parameter of this $\kappa$--symmetry depends on the boundary fermions which implies, as noticed already in \cite{Howe+Linstrom+Linus=2007}, that quantization of boundary fermions should result in an action possessing a  non-Abelian $\kappa$--symmetry. The previous attempts to construct the models with non-Abelian $\kappa$--symmetry gave negative results \cite{Bergshoeff:2001dc}.
Actually, this requirement of non-Abelian $\kappa$--symmetry comes from the fact that all the coordinate functions in the approach of \cite{Howe+Linstrom+Linus,Howe+Linstrom+Linus=2007} depend on the boundary fermions  so that, after quantization, all the coordinate functions become matrices and, to remove the extra unwanted $(p+1)$ bosonic and $16$ fermionic components one needs to have the reparametrization  and $\kappa$--symmetry with matrix parameters.

The problem with non-Abelian $\kappa$--symmetry appears at $(\alpha^\prime)^4$ order \cite{Bergshoeff:2001dc}. Probably, the further development of the boundary fermion approach will help to resolve it. However, even if it were confirmed  that the non-Abelian $\kappa$--symmetric DBI action is impossible to construct using the natural multibrane degrees of freedom, this would not imply that the approach of \cite{Howe+Linstrom+Linus=2007} is incorrect. It certainly provides a complete {\it classical} description of string theory with D-branes (or, better, 'pre-classical', see below). However, the consequent quantization of such a model implies simultaneous quantization of both the boundary fermions and coordinate functions. This would result in an appearance of not only the D$p$-brane worldvolume fields, but also of the bulk supergravity fields  and massive string state. A search for a Myers--like non-Abelian DBI-like action in this perspective  is reformulated as search  for a way to quantize {\it only} the boundary fermions leaving the classical description of the branes by coordinate functions untouched. Even if it happened that such a description is impossible to realize in its complete form, this could not be treated as incorrectness of the boundary fermion approach \cite{Howe+Linstrom+Linus=2007}, which gives a complete description of string theory D$p$--branes, but on the 'minus one quantized' level (considering the standard description of single D$p$-brane to be classical).

In our more traditional, but probably approximate, superembedding approach description of multiple brane systems only  the coordinate (super)fields corresponding to the center of mass motion are transformed by the target space Lorentz group transformations and $32$ component target space supersymmetry. The relative motion of constituent branes is described by the $SU(N)$  SYM multiplet, involving in addition to $d=(p+1)$ dimensional gauge potentials, only $(9-p)$  $su(N)$--valued matrix scalars ${\bb X}^i$ the Grassmann derivative of which is expressed through the $16$ fermionic  $su(N)$--valued matrix spinors $\Psi_\alpha$. The leading components of the superfields ${\bb X}^i$ and $\Psi_\alpha$ correspond to a {\it non-Abelian generalization of the static gauge coordinate functions} so that neither non-Abelian reparametrization invariance nor non-Abelian $\kappa$--symmetry is needed to reach the balance of degrees of freedom characteristic for a supersymmetric theory.

To conclude,
the existence of supersymmetric deformations of the SYM constraints in flat target superspace suggests that our choice of basic equations, including the superembedding equation and the constraints on the worldvolume $SU(N)$ SYM field strength, might be not unique also for the case of curved worldvolume superspace of a D--brane moving in a nontrivial supergravity background. However, we hope that even in this case, an approximate description given by our superembedding approach, corresponding to a low energy of relative motion and of the non-Abelian gauge field corresponding to it, but unrestricted (in the frame of DBI approximation) nonlinear description of the  $U(1)$ gauge fields and coordinate functions corresponding to the center of mass motion,  can be useful in future development of the fields.

Such a description in the frame of superembedding approach  has been shown to be allowed for  $p=0$ case, {\it i.e.} for the multiple D$p$-brane systems. An important problem is to check whether such a description is possible for higher branes. It is natural to begin with the simplest cases of multiple type IIB D$1$ and type IIA D$2$-brane systems. If the answer for  the second case happens to be affirmative, one can also search for similar superembedding description for the nearly coincident M$2$ branes which, if exists, should be related with the Bagger-Lambert--Gustavsson \cite{Bagger:2007jr}
and Aharony--Bergman--Jafferis--Maldacena \cite{Aharony:2008ug} models.

\section*{Acknowledgments}

The author thanks Dmitri Sorokin, Ulf Lindst\"om  and Linus Wulff for useful discussions on different stages of this work which was partially supported by the research grants FIS2008-1980 from the Spanish MICINN and grant
FIS2008-1980 and 38/50--2008 from the Ukrainian National
Academy of Sciences and Russian Federation RFFI.

\section*{Appendix A. Convenient representations for $11D$ Dirac matrices}
\renewcommand{\theequation}{A.\arabic{equation}}
\setcounter{equation}{0} \setcounter{subsection}{0}

\subsection*{A1. $SO(1,2)\otimes SO(8)$ invariant representation for $D=11$ Dirac matrices}

In superembedding description for M2-brane we use the following $SO(1,2)\otimes SO(8)$ invariant
representations for the eleven dimensional gamma matrices and charge conjugation matrix
\begin{eqnarray}\label{11DG=3+8}
& (\Gamma^{\underline{a}})_{\underline\alpha}{}^{\underline{\beta}}
 \equiv \left(\Gamma^{a}, \Gamma ^{i} \right)\; , \qquad a=0,1,2 \; , \qquad i=1,\ldots, 8 \; , \qquad
\nonumber \\  \label{11DG=3} & (\Gamma ^{a})_{\underline\alpha}{}^{\underline{\beta}}
\equiv \left(\Gamma ^{\underline{ 0}},~\Gamma^{\underline{ 9}}, {}~\Gamma^{\underline{
10}} \right) \equiv \left(\Gamma ^{{ 0}},\Gamma ^{{ 1}},\Gamma ^{{ 2}}\right) = \left(
\matrix{\gamma ^{a~ \beta}_{~\alpha} \delta _{qp} & 0 \cr 0 &  -
\gamma^{a}{}_{\beta}{}^{\alpha} \delta _{\dot{q} \dot{p}} }\right), \nonumber \\ \label{11DG=8}
& (\Gamma ^{i})_{\underline\alpha}{}^{\underline{\beta}} \equiv \left(\Gamma
^{1},\ldots , \Gamma ^{8}\right)= \left( \matrix{ 0 & -i\epsilon _{\alpha \beta} \gamma
^{i}_{{q}\dot{p}} \cr - i \epsilon ^{\alpha \beta} \tilde{\gamma}^{i}_{{\dot q}{p}}
\cr} \right) \; . \nonumber \\
\label{11DC=3+8}
& C^{\underline{\alpha}\underline{\beta}}= - C^{\underline{\beta}\underline{\alpha}}=
{\it diag} \left(i\epsilon ^{\alpha \beta}\delta_{qp},~ i \epsilon_{\alpha \beta}
\delta_{\dot{q}\dot{p}}\right), \quad C_{\underline{\alpha}\underline{\beta}} = \hbox{ {\it diag}} \left( -i \epsilon
_{\alpha \beta} \delta _{qp} ,~ -i \epsilon ^{\alpha\beta} \delta_{\dot{q}\dot{p}}
\right)
\; . \quad
\end{eqnarray}
Here $\gamma^a{}_\alpha{}^{\beta}$
and $\gamma ^{i}_{q\dot{q}}$ are the $SO(1,2)$ Dirac matrices and $SO(8)$
Pauli--matrices (Klebsh-Gordan coefficients), respectively. Some of  their useful properties are
\begin{eqnarray}
 \label{3Dgammas}
\gamma^a_{\alpha\beta}:= - i \gamma^a{}_\alpha{}^\gamma \epsilon_{\gamma\beta} =
\gamma^a_{\beta\alpha}=\gamma^a_{(\alpha\beta)}\; , \quad \tilde{\gamma}_a^{\alpha\beta}:= i \epsilon^{\alpha\gamma}  \gamma^a{}_\gamma{}^\beta= \tilde{\gamma}_a^{(\alpha\beta)} \; , \quad \epsilon^{\alpha\gamma}\epsilon_{\gamma\beta}=
\delta^{\alpha}_{\beta}\; , \qquad  \nonumber \\
\gamma^{ab}=-i\epsilon^{abc}\gamma_c \; , \quad
\gamma^a_{\alpha\beta}\tilde{\gamma}_a^{\gamma\delta}= 2 \delta_{(\alpha}{}^{\gamma}
\delta_{\beta )}{}^{\delta}
\; , \qquad \\
 \label{8Dgammas}
\tilde{\gamma}^i_{\dot{p} q}:= \gamma^i_{q\dot p} \; , \qquad \gamma^i_{q\dot
p}\gamma^j_{q \dot p} + \gamma^j_{q\dot p}\gamma^i_{q \dot p}= 2\delta^{ij}\delta_{qp}
\; , \qquad \gamma^i_{p\dot q}\gamma^j_{p \dot p} + \gamma^j_{p\dot q}\gamma^i_{p \dot
p}=
2\delta^{ij}\delta_{\dot{q}\dot{p}} \; , \qquad  \nonumber \\
\gamma^i_{q\dot q}\gamma^i_{p \dot p} = \delta_{qp} \delta_{\dot{q}\dot{p}}+ {1\over
4} \gamma^{ij}_{qp} \tilde{\gamma}^{ij}_{\dot{q}\dot{p}}  \qquad \Rightarrow  \qquad
\gamma^i_{(q|\dot q}\gamma^i_{|p) \dot p} = \delta_{qp} \delta_{\dot{q}\dot{p}} =
\gamma^i_{q(\dot q|}\gamma^i_{p |\dot p )}\; . \qquad
\end{eqnarray}
Notice that both 11D and 3d Dirac metrices are imaginary in our mostly minus signature conventions,
\begin{eqnarray}\label{11Deta=+--}
\eta^{\underline{a}\underline{b}}=diag(+,-,\ldots ,-)
\; , \qquad \eta^{{a}{b}}=diag(+,-,-)
\; . \qquad
\end{eqnarray}

Now we are ready to specify the relations (\ref{VGVT=uHa}) and (\ref{VTGV=uHa}) for the spinor moving frame
variables adapted to the (super)embedding of the M2-brane worldvolume (superspace):
\begin{eqnarray}\label{11Dug=v-Gv-AP}
 \delta_{qp} \tilde{\gamma}_a^{\alpha \beta} u^{~a}_{\underline{a}} =
v^{\alpha q} \tilde{\Gamma}_{\underline{a}} v^{\beta p} = -
v^{\alpha q} {\Gamma}_{\underline{a}} v^{\beta p} \; , \qquad \\
 \label{11Dug=v+Gv+AP} \delta_{\dot{q} \dot{p}}
{\gamma}_{a\alpha \beta} u^{a}_{\underline{a}} = v_{\alpha \dot q}
\tilde{\Gamma}_{\underline{a}} v_{\beta \dot p}= - v_{\alpha \dot q}
{\Gamma}_{\underline{a}} v_{\beta \dot p} \; ,
\qquad \\
 \label{11Dug=v+Gv-AP} \delta^{\alpha}_{\beta} \gamma ^{i}_{q
\dot p} u^{i}_{\underline{a}} = - v^{\alpha q}
\tilde{\Gamma}_{\underline{a}} v_{\beta \dot p} =   v^{\alpha q}
{\Gamma}_{\underline{a}} v_{\beta \dot p} \; , \qquad
\end{eqnarray}
and
\begin{eqnarray}
 \label{11DuaG=AP}
 u^{~a}_{\underline{b}} \Gamma^{\underline{b}}_{\underline{\alpha}\underline{\beta}} =
v_{\underline{\alpha}}^{\; \alpha q} (\gamma_a )_{\alpha \beta}
v_{\underline{\beta}}^{\; \beta q} + v_{\underline{\alpha} \alpha
\dot{q}} (\gamma_a )_{\alpha \beta} v_{\underline{\beta} \beta
\dot{q}} \; , \qquad \\
 \label{11DuiG=AP}
u^{~i}_{\underline{b}} \Gamma^{\underline{
b}}_{\underline{\alpha}\underline{\beta}} = -2
v_{(\underline{\alpha}|}{}^{\alpha q} \gamma^i_{q \dot{q}}
v_{|\underline{\beta})\; \alpha \dot{q}} \; . \qquad
\end{eqnarray}
An equivalent form for the set of relations (\ref{11DuaG=AP}), (\ref{11DuiG=AP}) is
\begin{eqnarray}
 \label{11DuaG=AP}
 u^{~a}_{\underline{b}} \tilde{\Gamma}{}^{\underline{b}\underline{\alpha}\underline{\kappa}} =
- v_{\alpha q}{}^{\underline{\alpha}} \, \tilde{\gamma}^{a \alpha \beta}
v_{\beta q}{}^{\underline{\kappa}} - v^{ \alpha
\dot{q}\, \underline{\alpha}} \, \gamma^a_{ \alpha \beta} v^{\beta
\dot{q}\, \underline{\kappa}} \; , \qquad \\
 \label{11DuiG=AP}
u^{~i}_{\underline{b}} \tilde{\Gamma}{}^{\underline{
b}\, \underline{\alpha}\underline{\beta}} = 2
v^{\alpha q\, (\underline{\alpha}|} \gamma^i_{q \dot{q}}
v_{\alpha \dot{q}}{}^{|\underline{\beta})} \; . \qquad
\end{eqnarray}

Another useful equation is the following `unity decomposition'
\begin{eqnarray}\label{I=vv+M2}
\delta_{\underline{\beta}}{}^{\underline{\alpha}} = i(v_{\underline{\beta}}{}^{\alpha q} v_{\alpha
q}{}^{\underline{\alpha}} +  v_{\underline{\beta} \alpha \dot q}
v^{\underline{\alpha} \alpha \dot q}) \; , \qquad
 \qquad
\end{eqnarray}
The difference of the contractions of the  same rank 16 blocks, $v_{\underline{\beta}}{}^{\alpha q} v_{\alpha
q}{}^{\underline{\alpha}} -  v_{\underline{\beta} \alpha \dot q}
v^{\underline{\alpha} \alpha \dot q}$, defines the matrix
\begin{eqnarray}\label{bG:=M2}
{\bar{\Gamma}}^{\underline{\alpha}}{}_{\underline{\beta}} := {i\over
3!} \varepsilon_{abc}({\Gamma} u^au^bu^c)^\alpha{}_\beta = {i\over
3!} \varepsilon_{abc} (v_{\alpha q}{}^{\underline{\alpha}}
\tilde{\gamma}^{abc \, \alpha \beta} v_{\beta
q}{}^{\underline{\gamma}} +
 v^{ \alpha \dot{q}\, \underline{\alpha}} \gamma^{abc}_{\alpha \beta}
v^{ \beta \dot{q}\, \underline{\gamma}})
C_{\underline{\gamma}\underline{\beta}}\; = \; \qquad \nonumber \\
= - i (v_{\underline{\beta}}{}^{\alpha q} v_{\alpha
q}{}^{\underline{\alpha}} -  v_{\underline{\beta} \alpha \dot q}
v^{\underline{\alpha} \alpha \dot q})\;  \qquad
 \qquad
\end{eqnarray}
entering the $\kappa$--symmetry projector of the standard formulation of M2-brane \cite{BST87}.

\subsection*{A2. $SO(1,5)\otimes SO(5)$ invariant representation for $D=11$ Dirac matrices}

In the superembedding description of M5-brane we use the following $SO(1,5)\otimes SO(5)$ invariant
representations for the eleven dimensional gamma matrices and charge conjugation matrix
\begin{eqnarray}\label{11DG=6+5}
& (\Gamma^{\underline{a}})_{\underline\alpha}{}^{\underline{\beta}}
 \equiv \left(\Gamma^{a}, \Gamma ^{i} \right)\; , \qquad a=0,1,\ldots,5 \; , \qquad i=1,\ldots, 5 \; , \qquad
\nonumber
\\
 \label{11DG=6} & (\Gamma ^{a})_{\underline\alpha}{}^{\underline{\beta}}
 = \left( \matrix{ 0 & -i\gamma^a_{\alpha \beta} \delta_q{}^p
 \cr +i\tilde{\gamma}{}^{a\, \alpha \beta} \delta_q{}^p
\cr} \right)\;  , \nonumber
 \\
  \label{11DG=5}
& (\Gamma ^{i})_{\underline\alpha}{}^{\underline{\beta}} \equiv \left(\Gamma
^{1},\ldots , \Gamma ^{8}\right)= \left(
\matrix{(\gamma ^iC)_q{}^p \delta_{\alpha}{}^{\beta} & 0 \cr 0 &  -(\gamma ^iC)_q{}^p \delta^{\alpha}{}_{\beta} }\right)
 \nonumber \\
\label{11DC=6+5}
& C^{\underline{\alpha}\underline{\beta}}= - C^{\underline{\beta}\underline{\alpha}}=
 \left( \matrix{ 0 & -i\delta^{\alpha}{}_{\beta} C
^{{q}{p}} \cr -i\delta_{\alpha}{}^{\beta} C
^{{q}{p}} & 0
\cr} \right)  \; , \quad C_{\underline{\alpha}\underline{\beta}}=
 \left( \matrix{ 0 & i\delta_{\alpha}{}^{\beta}  C
^{{q}{p}} \cr i\delta^{\alpha}{}_{\beta} C
^{{q}{p}} & 0
\cr} \right)  \;  . \quad
\end{eqnarray}
The SO(1,5) generalized Pauli matrices ($SU^*(4)$ Klebsh Gordan coefficients) are antisymmetric $\gamma^{a}_{\alpha\beta}=-\gamma^{a}_{\beta\alpha}=\gamma^{a}_{[\alpha\beta]}$, $\tilde{\gamma}_{a}^{\alpha\beta}=-\tilde{\gamma}_{a}^{\beta\alpha}=\tilde{\gamma}_{a}^{[\alpha\beta]}$ and possesses the following properties
\begin{eqnarray}
 \label{6Dgammas}
(\gamma^{(a}\tilde{\gamma}{}^{b )})_{\alpha}{}^{\beta} = \eta^{ab}\delta_{\alpha}{}^{\beta}\; , \qquad \eta^{ab}= diag (+,-,-,-,-,-)\; , \qquad \tilde{\gamma}_{a}^{\alpha\beta} ={1\over 2}\epsilon^{\alpha\beta\gamma\delta}{\gamma}_{a \gamma\delta}
\nonumber \\  \gamma^a_{\alpha\beta}\tilde{\gamma}_a^{\gamma\delta}= -4 \delta_{[\alpha}{}^{\gamma}
\delta_{\beta ]}{}^{\delta} \; , \qquad \gamma^a_{\alpha\beta} {\gamma}_{a \gamma\delta}= -2 \epsilon_{\alpha\beta\gamma\delta} \; , \quad  \gamma^{abcdef}{}_{\alpha}{}^{\beta}= \epsilon^{abcdef}\delta_{\alpha}{}^{\beta}  \nonumber \\
\gamma^{abc}{}_{\alpha\beta}=\gamma^{abc}{}_{(\alpha\beta )}= - {1\over 3!}\epsilon^{abcdef}\gamma_{def\;\alpha\beta}
\; , \qquad \tilde{\gamma}^{abc \alpha\beta}=\tilde{\gamma}^{abc (\alpha\beta )}=  {1\over 3!}\epsilon^{abcdef}\gamma_{def}^{\alpha\beta}
\; . \qquad
\end{eqnarray}
They are pseudo-real in the sense that the conjugate matrices $\gamma^{a *}_{\dot{\alpha}\dot{\beta}}:= (\gamma^{a}_{{\alpha}{\beta}})^*$ are expressed through $\gamma^{a}_{{\alpha}{\beta}}$ with the use of a matrix $B_\alpha{}^{\dot{\alpha}}$ \cite{PKT+6d} obeying $BB^*=-I$,
\begin{eqnarray}
 \label{6dG*=BGB}
(B\gamma^{a *}B^T):= B_\alpha{}^{\dot{\alpha}}
\gamma^{a *}_{\dot{\alpha}\dot{\beta}} B_\beta{}^{\dot{\beta}}= \gamma^{a}_{{\alpha}{\beta}}
\; , \qquad (B^{*T}\tilde{\gamma}^{a *}B^{*}):= B^*{}_{\dot{\alpha}}{}^\alpha
\tilde{\gamma}{}^{a * \dot{\alpha}\dot{\beta}} B_\beta{}^{\dot{\beta}}=
\gamma^{a}_{{\alpha}{\beta}}
\; , \qquad \\ \nonumber
B_\alpha{}^{\dot{\beta}}B^*{}_{\dot{\beta}}{}^{\beta}= - \delta_\alpha{}^{\beta}\; . \qquad
\end{eqnarray}

The properties of the $SO(5)$ Klebsh-Gordan coefficients (generlizaed
Pauli matrices) and of the charge conjugation matrix are
\begin{eqnarray}\label{sigma5d=} {\gamma}^{i}_{qp}=-
{\gamma}^{i}_{pq}= - (\tilde{\gamma}^{i qp})^*=  {1\over 2}\epsilon_{qprs}
\tilde{{\gamma}}^{i\, rs} \; ,  \qquad i=1,\ldots, 5\; , \quad  q,p,r,s=1,\ldots , 4\; , \qquad \nonumber \\
\label{Cliff5d} \fbox{\, ${\gamma}^{i}\tilde{\gamma}^{j}+
{\gamma}^{j}\tilde{{\gamma}}^{i}= 2{\delta}^{ij} \delta_q{}^p$\,} \; ,
\qquad {\gamma}^{i}\tilde{\gamma}^{j}- {\gamma}^{j}\tilde{{\gamma}}^{i}=:
2{{\gamma}}^{ij}{}_q{}^p\; , \qquad  \\
\label{5d=C} C_{qp}=-
C_{pq}= - (C^{qp})^* =  {1\over 2}\epsilon_{qprs}
C^{rs} \; ,  \qquad  C_{qr}C^{rp}=\delta_q{}^p\; , \qquad \nonumber \\   C\tilde{\gamma}^iC = - {\gamma}^i \; ,  \qquad C{\gamma}^iC = - \tilde{\gamma}^i \; , \qquad   \\
\label{so(5)id} {\gamma}^{i}_{qp}\tilde{\gamma}^{i rs}= -4
\delta_{[q}{}^{r}\delta_{p]}{}^{s}- C_{qp}C^{rs}\; , \qquad {\gamma}^{i}_{qp}\,
{\gamma}^{i}_{rs} = -2\epsilon_{qprs} - C_{qp}C_{rs} \; . \qquad
\end{eqnarray}

These properties can be deduced from the properties of $SU(4)$ Klebsh-Gordan coefficients ${\gamma}^{I}_{ji}$, $\tilde{{\gamma}}^{I\, ij}$, $I=1,\ldots, 6$,
\begin{eqnarray}\label{sigma6d=}
{\gamma}^{I}_{qp}= - {\gamma}^{I}_{pq}= - (\tilde{\gamma}^{I qp})^*=  {1\over 2}\epsilon_{pqrs}
\tilde{{\gamma}}^{I\, rs} \; ,  \qquad I=1,\ldots, 6\; , \quad  p,q,r,s=1,\ldots , 4\; , \qquad \\
\label{Cliff6d} \fbox{\, ${\gamma}^{I}\tilde{\gamma}^{J}+
{\gamma}^{J}\tilde{{\gamma}}^{I}= 2{\delta}^{IJ} \delta_q{}^p$\,} \; ,
\qquad {\gamma}^{I}\tilde{\gamma}^{J}- {\gamma}^{J}\tilde{{\gamma}}^{I}=:
2{{\gamma}}^{IJ}{}_q{}^p\; , \qquad  \\
\label{so(6)id} {\gamma}^{I}_{qp}\tilde{\gamma}^{I rs}= -4
\delta_{[q}{}^{r}\delta_{p]}{}^{s}\; , \qquad {\gamma}^{I}_{qp}\,
{\gamma}^{I}_{rs} = -2\epsilon_{qprs}\; , \qquad
\end{eqnarray}
after identification ${\gamma}^{I}_{qp}= ({\gamma}^{i}_{qp}, C_{qp})$, $i=1,...,5$.

\bigskip

The relations (\ref{VGVT=uHa}) and (\ref{VTGV=uHa}) for the spinor moving frame
variables adapted to the (super)embedding of the M5-brane worldvolume superspace are given by
\begin{eqnarray}\label{M5:v1Gv1=}
 & v^{\alpha q} \tilde{\Gamma}_{\underline{a}} v^{\beta p} = \; u_{\underline{a}}{}^b \tilde{\gamma}_b{}^{\alpha \beta} C^{qp}\; , \qquad & v_{\alpha q} {\Gamma}_{\underline{a}} v_{\beta p} = \; u_{\underline{a}}{}^b {\gamma}_{b\, \alpha \beta} C_{qp}\; , \qquad
 \\ \label{M5:v2Gv2=}
 & v_{\alpha}{}^{q} \tilde{\Gamma}_{\underline{a}} v_{\beta}{}^{p} = - u_{\underline{a}}{}^b {\gamma}_{b\, \alpha \beta} C_{qp}\; , \qquad & v_{q}{}^{\alpha} {\Gamma}_{\underline{a}} v_{p}{}^{\beta} = -u_{\underline{a}}{}^b \tilde{\gamma}_b{}^{\alpha \beta} C_{qp}\; , \qquad
\\ \label{M5:v1Gv2=} &
v^{\alpha q} \tilde{\Gamma}_{\underline{a}} v_{\beta}^{p} = \; i u_{\underline{a}}{}^i \tilde{\gamma}^i {}^{qp}\delta^{\alpha}{}_{\beta}\; , \qquad & v_{\alpha q} {\Gamma}_{\underline{a}} v^{\beta}_{p} = -i u_{\underline{a}}{}^i {\gamma}^i{}_{qp}\delta_{\alpha}{}^{\beta}\; , \qquad
\end{eqnarray}
and
\begin{eqnarray}
 \label{M5:uaG=}
 u^{~a}_{\underline{b}} \Gamma^{\underline{b}}_{\underline{\alpha}\underline{\beta}} =
v_{\underline{\alpha}}^{\; \alpha q} (\gamma_a )_{\alpha \beta}
v_{\underline{\beta}}^{\; \beta p} C_{qp}- v_{\underline{\alpha} \alpha
}{}^q \tilde{\gamma}_a^{\alpha \beta} v_{\underline{\beta} \beta}{}^p C_{qp} \; , \qquad \\
 \label{M5:uiG=}
u^{~i}_{\underline{b}} \Gamma^{\underline{
b}}_{\underline{\alpha}\underline{\beta}} = 2i
v_{(\underline{\alpha}|}{}^{\alpha q} \gamma^i_{q p}
v_{|\underline{\beta})\alpha}{}^p \; , \qquad
\end{eqnarray}
as well as
\begin{eqnarray}
 \label{M5:uatG=}
 u^{~a}_{\underline{b}} \tilde{\Gamma}{}^{\underline{b}\underline{\alpha}\underline{\beta}} =
v_{\alpha q}{}^{\underline{\alpha}} \, \tilde{\gamma}^{a \alpha \beta}
v_{\beta p}{}^{\underline{\beta}} C^{qp} - v_q{}^{\alpha
 \underline{\alpha}} \, \gamma^a_{ \alpha \beta} v_p^{\beta
\underline{\beta}} C^{qp}\; , \qquad \\
 \label{M5:uitG=}
u^{~i}_{\underline{b}} \tilde{\Gamma}{}^{\underline{
b}\, \underline{\alpha}\underline{\beta}} = -2i
v_{\alpha q}^{\; (\underline{\alpha}|} \tilde{\gamma}^{i \,qp}
v_p{}^{\alpha|\underline{\beta})} \; . \qquad
\end{eqnarray}
The `unity decomposition' reads simply as
\begin{eqnarray}\label{I=vv+M5}
\delta_{\underline{\beta}}{}^{\underline{\alpha}} = v_{\underline{\beta}}{}^{\alpha q} v_{\alpha
q}{}^{\underline{\alpha}} +  v_{\underline{\beta} \alpha}{}^q
v_q^{ \alpha\underline{\alpha}} \; , \qquad
 \qquad
\end{eqnarray}
but the components of the inverse spinor moving frame matrix, $ v_{\alpha
q}{}^{\underline{\alpha}}$ and  $v_q^{ \alpha\underline{\alpha}}$  are expressed through  $v_{\underline{\beta}}{}^{\alpha q} $ and $v_{\underline{\beta} \alpha}{}^q$ by
(\ref{v-1=CvM5}).

The derivatives of spinor moving frame variables are expressed through the generalized Cartan forms by   \begin{eqnarray}\label{Dvaq=}
& Dv_{\underline{\alpha}}{}^{\alpha q}= {i\over 2}\Omega^{ai}  v_{\underline{\alpha} \beta}{}^p
\tilde{\gamma}_a^{\beta\alpha} ({\gamma}^iC)_p{}^q \; , \qquad & Dv_{\alpha q}{}^{\underline{\alpha}}= {i\over 2}\Omega^{ai}
{\gamma}_{a \, \alpha\beta} ({\gamma}^iC)_q{}^p v_p^{\beta\underline{\alpha}}\; , \qquad
 \qquad \\ \label{Dvaaq=} & Dv_{\underline{\alpha}\alpha }{}^{q}= - {i\over 2}\Omega^{ai}  v_{\underline{\alpha}}^{\beta p}
\tilde{\gamma}_{a\, \beta\alpha} ({\gamma}^iC)_p{}^q \; , \qquad & Dv_q{}^{\alpha\underline{\alpha}}= - {i\over 2}\Omega^{ai}\tilde{\gamma}_a^{\alpha\beta}  ({\gamma}^iC)_q{}^p v_{\beta p}{}^{\underline{\alpha}}\; . \qquad
 \qquad
\end{eqnarray}

\bigskip

\section*{
Appendix B. Some details on type IIA supergravity superspace.}
\renewcommand{\theequation}{B.\arabic{equation}}
\setcounter{equation}{0} \setcounter{subsection}{0}

\bigskip

The type IIA superspace geometry was worked out in \cite{Bellucci+Gates+89}. Here we present some equations in our present notation.

Fermionic torsion of general type IIA supergravity superspaces reads
\begin{eqnarray}\label{Tal1=str}
& T^{\alpha 1} =  {1\over 2}E^{\underline{b}} \wedge
E^{\underline{a}}
 T_{\underline{a}\underline{b}}{}^{\alpha 1 }  +
E^{\underline{a}}\wedge {\cal E}^{{\underline{\beta}}}
T_{{\underline{\beta}}\underline{a}}{}^{\alpha 1 }
 - 2i E^{\alpha 1}\wedge E^{\beta 1} \Lambda_{\beta 1} + i E^{1}\sigma^{\underline{a}}\wedge E^{1}\,
\tilde{\sigma}_{\underline{a}}^{\alpha\beta} \Lambda_{\beta
1}  \, ,  \, \nonumber \\
\label{Tal2=str}
 & T_{\alpha}^{ 2} =
{1\over 2}E^{\underline{b}} \wedge E^{\underline{a}}
 T_{\underline{a}\underline{b}}{}_{\alpha}^{ 2} +
E^{\underline{a}}\wedge {\cal E}^{{\underline{\beta}}}
T_{{\underline{\beta}}\underline{a}}{}_{\alpha}^{ 2}
 -   2i  E^2_{\alpha}\,\wedge
E^2_{\beta} \; \Lambda_2^{\beta}\,  + i
E^{2}\tilde{\sigma}_{\underline{a}}\wedge E^{2}\,
{\sigma}^{\underline{a}}_{\alpha\beta} \, \Lambda_2^{\beta }  \; ,
\qquad
\end{eqnarray}
where
\begin{eqnarray}\label{2iLambda=DP}
 {\cal E}^{{\underline{\beta}}}=(E^{\beta 1}, E_{\beta}{}^{2})\; , \qquad
\Lambda_{\beta 1} ={i\over 2}D_{\beta 1}\Phi \; , \qquad \Lambda^{\beta}{}_{2}={i\over 2}D^{\beta}{}_{2}\Phi\; , \qquad
\end{eqnarray} and
\begin{eqnarray}\label{T1b1=IIA}
&& T_{\beta 1\, \underline{a}\;}{}^{\gamma 1} =  - {1\over 8}
H_{\underline{a}\underline{b}\underline{c}}
(\sigma^{\underline{b}\underline{c}})_{\beta}{}^{\gamma} =
T^{\,\gamma}_{\; 2\; \underline{a}\;}{}_{\beta 1} \; ,  \qquad
\\ \label{T1b2=IIA}  T_{\beta 1\, \underline{a}\;}{}_{\gamma}{}^{ 2}
&=&
 {e^{\Phi}\over 8\cdot 2!}
R_{\underline{b}\underline{c}}
(\sigma_{\underline{a}}\tilde{\sigma}^{\underline{b}\underline{c}})_{\beta\gamma}
- {e^{\Phi}\over 8\cdot 4!}
R_{\underline{b}\underline{c}\underline{d}\underline{e}}
(\sigma_{\underline{a}}\tilde{\sigma}^{\underline{b}\underline{c}\underline{d}\underline{e}})_{\beta\gamma}
+ \qquad \nonumber  \\ && +  {i\over 8}
\Lambda_2\sigma_{\underline{b}\underline{c}}\Lambda_1\;
(\sigma_{\underline{a}}\tilde{\sigma}^{\underline{b}\underline{c}})_{\beta\gamma}
- {3i\over 16\cdot 4!}
\Lambda_2\sigma_{\underline{b}\underline{c}\underline{d}\underline{e}}\Lambda_1\;
(\sigma_{\underline{a}}\tilde{\sigma}^{\underline{b}\underline{c}\underline{d}\underline{e}})_{\beta\gamma}
\; ,  \\
\label{T2b1=IIA}  T^{\,\beta}_{\; 2\; \underline{a}\;}{}^{\gamma\,
1} &=&  {e^{\Phi}\over 8\cdot 2!} R_{\underline{b}\underline{c}}
(\tilde{\sigma}_{\underline{a}}{\sigma}^{\underline{b}\underline{c}})^{\beta\gamma}
+ {e^{\Phi}\over 8\cdot 4!}
R_{\underline{b}\underline{c}\underline{d}\underline{e}}
(\tilde{\sigma}_{\underline{a}}{\sigma}^{\underline{b}\underline{c}\underline{d}\underline{e}})^{\beta\gamma}
+  \qquad \nonumber \\ && +  {i\over 8}
\Lambda_2\sigma_{\underline{b}\underline{c}}\Lambda_1\;
(\tilde{\sigma}_{\underline{a}}{\sigma}^{\underline{b}\underline{c}})^{\beta\gamma}
+ {3i\over 16\cdot 4!}
\Lambda_2\sigma_{\underline{b}\underline{c}\underline{d}\underline{e}}\Lambda_1\;
(\tilde{\sigma}_{\underline{a}}{\sigma}^{\underline{b}\underline{c}\underline{d}\underline{e}})^{\beta\gamma}
\; \, . \;
\end{eqnarray}

The Riemann curvature 2-form of the type IIA superspace is expressed
through the above dim 1  torsion components and through the dim 3/2
ones by the solution of Bianchi identities,
\begin{eqnarray}\label{Rab=IIA}
{\bb R}^{\underline{a}\underline{b}}&:=&
d\omega^{\underline{a}\underline{b}} -
\omega^{\underline{a}\underline{c}}\wedge
\omega_{\underline{c}}{}^{\underline{b}} =  2i E^{\alpha 1}\wedge
E^{\beta 1} \sigma^{[\underline{a}}{}_{\gamma (\alpha}T_{\beta )
1}{}^{\underline{b}] \,\gamma 1} + 2i E^2_{\alpha }\wedge
E^2_{\beta} \tilde{\sigma}{}^{[\underline{a}\gamma
(\alpha}T_2^{\beta )}{}^{\underline{b}]}{}^2_{\gamma } + \qquad
\nonumber \\ && + 4i E^{\alpha 1}\wedge E^2_{\beta}
\left(\sigma^{[\underline{a}}{}_{\gamma \alpha} T_2^{\beta
}{}^{\underline{b}]}{}^{\gamma 1} +
\tilde{\sigma}{}^{[\underline{a}|\, \gamma \beta}T_{\alpha 1\; }
{}^{|\underline{b}]}{}^2_{\gamma }\right)+ \qquad \nonumber \\ &&  +
E^{\underline{c}}\wedge
 E^{\alpha 1} \left( 2iT_{\underline{c}}{}^{[\underline{a}|\, \beta 1}\sigma^{|\underline{b}]}{}_{\beta\alpha}
  - i T^{\underline{a}\underline{b}\, \beta 1}\sigma_{\underline{c}\,
  \beta\alpha}\right)+  \qquad \nonumber \\ &&  +
E^{\underline{c}}\wedge
 E^2_{\alpha} \left( 2iT_{\underline{c}}{}^{[\underline{a}|}{}^{ \, 2}_{\beta}\tilde{\sigma}{}^{|\underline{b}] \beta\alpha}
  - i T^{\underline{a}\underline{b}\,2}_{\beta}\sigma_{\underline{c}\,
  \beta\alpha}\right) + {1\over 2}E^{\underline{d}}\wedge E^{\underline{c}}\;  {\bb R}_{\underline{c}\underline{d}}{}^{\underline{a}\underline{b}}
\; . \qquad
\end{eqnarray}

The following equations are also useful
\begin{eqnarray}\label{D1L1=IIA}
D_{\alpha 1}\Lambda_{\beta 1}:= {i\over 2} D_{\alpha 1}D_{\beta
1}\Phi= - {1\over
2}\sigma^{\underline{a}}_{\alpha\beta}D_{\underline{a}}\Phi +
{1\over 4!}
 \left( H_{\underline{a}\underline{b}\underline{c}}- {i\over 2}\,  \Lambda_1
 \tilde{\sigma}_{\underline{a}\underline{b}\underline{c}}\Lambda_1\right)\;
 \sigma^{\underline{a}\underline{b}\underline{c}}_{\alpha\beta}\; , \qquad
\\ \label{D2L2=IIA}
D_2^{\alpha}\Lambda_2^{\beta} :={i\over 2}
D_2^{\alpha}D_2^{\beta}\Phi\, = - {1\over 2}
\tilde{\sigma}^{\underline{a}\, \alpha\beta}D_{\underline{a}}\Phi +
{1\over 4!}
 \left( H_{\underline{a}\underline{b}\underline{c}}+ {i\over 2}\,  \Lambda_2
 \sigma_{\underline{a}\underline{b}\underline{c}}\Lambda_2\right)
 \tilde{\sigma}^{\underline{a}\underline{b}\underline{c}\,\alpha\beta}\; , \qquad
\\  \label{D2L1=IIA}
D_2^{\beta} \Lambda_{\alpha 1}:={i\over 2} D_2^{\beta}D_{\alpha 1}
\Phi\, =   (t-i \Lambda_2\Lambda_1) \,\delta_{\alpha }{}^{\beta} +
 {3\over 16}
 \left( e^{\Phi}R_{\underline{a}\underline{b}}+ 2i\,  \Lambda_2
 \sigma_{\underline{a}\underline{b}}\Lambda_1\right)
 \tilde{\sigma}^{\underline{a}\underline{b}}{}_{\alpha }{}^{\beta}+ \qquad \nonumber \\ +
 {1\over 8\cdot 4!}
 \left( e^{\Phi}R_{\underline{a}\underline{b}\underline{c}\underline{d}}+ {3i\over 2}\,
 \Lambda_2
 \sigma_{\underline{a}\underline{b}\underline{c}\underline{d}}\Lambda_1\right)
 \tilde{\sigma}^{\underline{a}\underline{b}\underline{c}\underline{d}}{}_{\alpha }{}^{\beta}
 \qquad \nonumber \\ = - D_{\alpha 1} \Lambda_2^{\beta}  :=-{i\over 2}D_{\alpha 1}
D_2^{\beta}\Phi\, \; . \qquad
\end{eqnarray}

\section*{Appendix C. Structure of fermionic equations for multiple
D$0$-brane system in the presence of fluxes}

The fermionic equations of motion which follows from our superembedding description of  multiple D$0$-brane system  in general type IIA supergravity background
have the structure of \cite{IB09:D0}
\begin{eqnarray}\label{D0:Dirac=LR}
 {7\over 8}\left( D_{0}\Psi- {1\over 4} [ {\bb X}^i\, , \,
(\sigma^{0i}\Psi)]\right)&=&
(e^{\hat{\Phi}}\hat{R}{}^{0i}+\widehat{D^i\Phi})(\sigma^{0i}\Psi) + {1\over 64} \hat{H}^{0ij} \sigma^{0k}
\sigma^{ij} \sigma^{0k}\Psi -
\qquad \nonumber  \\ & -& {1\over 64} \sigma^{0k}\left(  - {1\over 2!}
e^{\hat{\Phi}}\hat{R}_{\underline{b}\underline{c}}
\sigma^{\underline{b}\underline{c}}-
    {1\over
4!}e^{\hat{\Phi}}\hat{R}_{\underline{b}\underline{c}\underline{d}\underline{e}}
\sigma^{\underline{b}\underline{c}\underline{d}\underline{e}}\right)
\sigma^{0k}\Psi + \qquad \nonumber \\ & +&   D_0{\bb
X}^i \left(a_1 \sigma^{0i}\hat{\Lambda}_{ 1}
+ a_2 \sigma^{i}\hat{\Lambda}_2\right) +
\qquad \nonumber \\ &+& {1\over 16} [{\bb
X}^i, {\bb X}^j] \left(b_1
\sigma^{ij} \hat{\Lambda}_{ 1} - b_2
\sigma^{0ij}\hat{\Lambda}_2\right)  + \qquad \nonumber \\ &+&
{\cal O}(\hat{\Lambda}_{1,2}\! \cdot \hat{\Lambda}_{1,2}\cdot \Psi)
\qquad
\end{eqnarray}
with some constants $a_{1,2}$ and $b_{1,2}$.


{\small
\begin{thebibliography}{99}

\bibitem{BST87} E. Bergshoeff, E. Sezgin and P.K. Townsend, {\it Supermembrane and
eleven-dimansional supergravity}, {\em Phys. Lett.} {\bf B189}, 75-78 (1987); {\em Ann.
Phys. (NY)} {\bf 185}, 330 (1988).

\bibitem{AETW} A. Ach\'ucarro, J.M. Evans, P.K. Townsend and D.L. Wiltshire,
{\it super-p-branes} 
{\em Phys. Lett.} {\bf B198}, 441 (1987).

\bibitem{hs96}
 P.~S.~Howe and E.~Sezgin,
  {\it Superbranes},
  Phys.\ Lett.\  B {\bf 390}, 133 (1997)
  [hep-th/9607227].


\bibitem{Dpac} M.\ Cederwall, A.\ von Gussich, B.E.W.\ Nilsson, A.\ Westerberg, {\it The
Dirichlet super-three-branes in ten-dimensional type IIB
supergravity},  {Nucl.Phys.} {\bf B490}
(1997) 163--178  [hep-th/9610148]; \\
M.\ Aganagic, C.\ Popescu, J.H.\ Schwarz, {\sl D-brane actions with
local kappa symmetry}, {Phys.Lett.} {\bf B393} (1997) 311--315,
[hep-th/9610249].


\bibitem{hs2}
  P.~S.~Howe and E.~Sezgin,
  {\it D = 11, p = 5},
  Phys.\ Lett.  {\bf B394}, 62 (1997)
  [hep-th/9611008].

\bibitem{Dpac2} M.~
Cederwall, A.\ von Gussich, B.E.W.\ Nilsson, P.\ Sundell and A.\
Westerberg, {\it The Dirichlet super-p-branes in ten-dimensional
type IIA and IIB supergravity}, {Nucl.Phys.} {\bf B490} (1997) 179--201
[hep-th/9611159];  M.\ Aganagic, C.\ Popescu, J.H.\ Schwarz,  {\it Gauge--invariant and
gauge--fixed D-brane actions}, {Nucl.Phys.} {\bf B490} (1997) 202,
[hep-th/9612080].

\bibitem{B+T=Dpac}
E.\ Bergshoeff, P.K.\ Townsend, {\it Super-$D$-branes}, {Nucl.Phys.}
{\bf B490} (1997) 145--162, [hep-th/9611173];

\bibitem{blnpst}
  I.~A.~Bandos, K.~Lechner, A.~Nurmagambetov, P.~Pasti, D.~P.~Sorokin and M.~Tonin,
  {\it Covariant action for the super-five-brane of M-theory},
  Phys.\ Rev.\ Lett.\  {\bf 78} (1997) 4332
  [hep-th/9701149].

\bibitem{schw5}
M.~Aganagic, J.~Park, C.~Popescu and J.~H.~Schwarz,
  {\it World-volume action of the M-theory five-brane},
  Nucl.\ Phys.\  {\bf B496}, 191-214 (1997)
  [hep-th/9701166].

\bibitem{Duality}
C.M.~Hull and P.~K.~Townsend, {\it Unity of superstring dualities}, Nucl.\ Phys.\ {\bf
B438}, 109-137 (1995) [hep-th/9410167]; E.~Witten, {\it String theory dynamics in
various dimensions}, Nucl.\ Phys.\  {\bf B443}, 85 -126 (1995) [hep-th/9503124].

\bibitem{M-theory}
J.H. Schwarz, {\it Second Superstring Revolution},
{Nucl.Phys.Proc.Suppl.} {\bf B55}, 1-32 (1997) (hep-th/9607067);
\\ P.K. Townsend, {\sl
Four Lectures on M--theory}, In :{\sl Summer School High energy physics and cosmology,
Trieste 1996} (Eds. E. Gava, A. Masiero, K.S. Narain, S. Randjbar-Daemi, Q. Shafi), The
ICTP Series in Theoretical Physics, Vol. 13, World Scientific, 1997, Singapore,
pp. 385-438 [hep-th/9612121].


\bibitem{AdS/CFT}
J.M. Maldacena, {Adv. Theor. Math. Phys.} {\bf 2}, 231-252 (1998) (hep-th/9711200); \,
S.S. Gubser, I.R. Klebanov and A.M. Polyakov, {\sl Phys.Lett.}
 {\bf B428}, 105-114 (1998) [{hep-th/9802109}], \, E. Witten, {Adv. Theor. Math. Phys.} {\bf 2}, 253-291 (1998)
[{hep-th/9802150}].

\bibitem{AdS/CFTrev}
O. Aharony, S.S. Gubser, J. Maldacena, H. Ooguri and Y. Oz,
``Large N field theories, string theory and gravity,''
  Phys.\ Rept.\  {\bf 323}, 183 (2000)
  [arXiv:hep-th/9905111];
  L.~F.~Alday and R.~Roiban,
  ``Scattering Amplitudes, Wilson Loops and the String/Gauge Theory
  Correspondence,''
  Phys.\ Rept.\  {\bf 468}, 153-211 (2008)
  [arXiv:0807.1889 [hep-th]];
  S.~S.~Gubser,
  ``Using string theory to study the quark-gluon plasma: progress and perils,''
  in {\it  Proceedings of Quark Matter 2009: 21st International Conference on Ultra-Relativistic Nucleus-Nucleus Collisions (QM2009), Knoxville, Tennessee, 30 Mar - 4 Apr 2009}, Nucl.\ Phys.\  A {\bf 830}, 657C (2009)
  [arXiv:0907.4808 [hep-th]] and refs. therein

\bibitem{SUGRA}
M.J. Duff, R.R. Khuri and J.X. Lu, ``{String Solitons},''
{\sl Phys. Rep.} {\bf 259}, 213-326 (1995); \\
K.S. Stelle, {BPS Branes in Supergravity}, hep-th/9803116.
in: {\it High-Energy Physics And Cosmology, Proc.  ICTP Summer School},
2 Jun - 11 Jul 1997, Trieste, Italy
(E. Gava, {\it et al.} Eds.)
A. Masiero, K.S. Narain, S. Randjbar-Daemi,
G.~Senjanovic, A. Smirnov, Q. Shafi),
World Scientific, Singapore, 1998, p. 29-127
(hep-th/9803116) {and refs. therein}.


\bibitem{bpstv}
  I.~A.~Bandos, D.~P.~Sorokin, M.~Tonin, P.~Pasti and D.~V.~Volkov,
  {\it Superstrings and supermembranes in the doubly supersymmetric geometrical
  approach},
  Nucl.\ Phys.\  {\bf B446}, 79-118 (1995)
  [hep-th/9501113].



\bibitem{bst97}
  I.~A.~Bandos, D.~P.~Sorokin and M.~Tonin,
  {\it Generalized action principle and superfield equations of motion for  $D = 10$
  D$p$-branes},
  Nucl.\ Phys.\  {\bf B497}, 275-296 (1997)
  [hep-th/9701127].

\bibitem{HS+Chu=PLB98}
  C.~S.~Chu, P.~S.~Howe and E.~Sezgin,
  {\it Strings and D-branes with boundaries},
  Phys.\ Lett.\  {\bf B428}, 59-67 (1998)
  [hep-th/9801202];
  C.~S.~Chu, P.~S.~Howe, E.~Sezgin and P.~C.~West,
  {\it Open superbranes},
  Phys.\ Lett.\  {\bf B429}, 273--280 (1998)
  [hep-th/9803041].


\bibitem{ABKZ}
V.~Akulov, I.~A.~Bandos, W.~Kummer and V.~Zima,
  {\it $D = 10$ Dirichlet super-$9$-brane},
  Nucl.\ Phys.\   {\bf B527}, 61 (1998)
  [hep-th/9802032].




\bibitem{Dima99}
  D.~P.~Sorokin,
  {\it Superbranes and superembeddings},
  Phys.\ Rept.\  {\bf 329}, 1 (2000)
  [hep-th/9906142].

\bibitem{B00}
I.~A.~Bandos,
  {\it Superembedding approach and S-duality: A unified description of
  superstring and super-D1-brane}, Nucl.\ Phys.\  {\bf B599}, 197 (2001)
  [arXiv:hep-th/0008249].



\bibitem{IB08:Q7}
  I.~A.~Bandos,
  {\it On superembedding approach to type IIB 7-branes},
  JHEP {\bf 0904}, 085 (2009)
  [arXiv:0812.2889 [hep-th]].

\bibitem{IB09:D0}
  I.~A.~Bandos,
  ``On superembedding approach to multiple D-brane system. D0 story,''
  Phys.\ Lett.\  B {\bf 680}, 267 (2009)
  [arXiv:0907.4681 [hep-th]].


\bibitem{Emparan:1997rt}
  R.~Emparan,
  {\it Born-Infeld strings tunneling to D-branes},
  Phys.\ Lett.\  B {\bf 423}, 71 (1998)
  [arXiv:hep-th/9711106].


\bibitem{Myers:1999ps}
  R.~C.~Myers,
  {\it Dielectric-branes},
  JHEP {\bf 9912}, 022 (1999)
  [hep-th/9910053].



\bibitem{Howe+Linstrom+Linus}
 P.~S.~Howe, U.~Lindstrom and L.~Wulff,
  {\it Superstrings with boundary fermions},
  JHEP {\bf 0508}, 041 (2005)
  [hep-th/0505067];
 {\it On the covariance of the Dirac-Born-Infeld-Myers action},
  JHEP {\bf 0702}, 070 (2007)
  [hep-th/0607156];



\bibitem{Howe+Linstrom+Linus=2007}
P.~S.~Howe, U.~Lindstrom and L.~Wulff,
  {\it Kappa-symmetry for coincident D-branes},
  JHEP {\bf 0709}, 010 (2007)
  [arXiv:0706.2494 [hep-th]].

\bibitem{Sagnotti}
 A.~Sagnotti, {\it Open strings and their symmetry groups}, in:
 {\it NATO Advanced Summer Institute on Nonperturbative Quantum Field
Theory (Cargese Summer Institute) Cargese, France, Jul 16-30, 1987},
G. 't Hooft, A. Jaffe, G. Mack, P.K. Mitter, R. Stora Eds., Plenum
Press, 1988, pp. 521-528  [Preprint
  ROM2F-87-25  now also available as
arXiv:hep-th/0208020].

\bibitem{Horava}
P.~Horava, {\it Strings on world sheet orbifolds},
  Nucl.\ Phys.\  {\bf B327}, 461 (1989);
  {\it Background duality of open string models},
  Phys.\ Lett.\  {\bf B231}, 251 (1989).

\bibitem{Polchinski89}
  J.~Dai, R.~G.~Leigh and J.~Polchinski,
  {\it New Connections Between String Theories},
  Mod.\ Phys.\ Lett.\  {\bf A4}, 2073 (1989).

\bibitem{Leigh:1989jq}
  R.~G.~Leigh,
  {\it Dirac-Born-Infeld Action from Dirichlet Sigma Model},
  Mod.\ Phys.\ Lett.\  A {\bf 4}, 2767 (1989).


\bibitem{Polchinski95}
  J.~Polchinski,
  {\it Dirichlet-Branes and Ramond-Ramond Charges},
  Phys.\ Rev.\ Lett.\  {\bf 75}, 4724 (1995)
  [hep-th/9510017].





\bibitem{Witten:1995im}
  E.~Witten,
  {\it Bound states of strings and p-branes},
  Nucl.\ Phys.\   {\bf B460}, 335 (1996)
  [hep-th/9510135].



\bibitem{D0-1996}
U.~H.~Danielsson, G.~Ferretti and B.~Sundborg,
  {\it D-particle Dynamics and Bound States},
  Int.\ J.\ Mod.\ Phys.\  {\bf A11}, 5463 (1996)
  [hep-th/9603081];
D.~N.~Kabat and P.~Pouliot,
  {\it A Comment on Zero-brane Quantum Mechanics},
  Phys.\ Rev.\ Lett.\  {\bf 77}, 1004-1007 (1996)
  [hep-th/9603127];\\
M.~R.~Douglas, D.~N.~Kabat, P.~Pouliot and S.~H.~Shenker,
  {\it D-branes and short distances in string theory},
  Nucl.\ Phys.\   {\bf B485}, 85-127 (1997)
  [hep-th/9608024].




\bibitem{Banks:1996vh}
  T.~Banks, W.~Fischler, S.~H.~Shenker and L.~Susskind,
  {\it M theory as a matrix model: A conjecture},
  Phys.\ Rev.\  {\bf D55}, 5112-5128 (1997)
  [hep-th/9610043].




\bibitem{Townsend95}
 P.~K.~Townsend,
  {\it D-branes from M-branes},
  Phys.\ Lett.\  {\bf B373}, 68-75 (1996)
  [hep-th/9512062].



\bibitem{D-braneAnom}
  M.~B.~Green, J.~A.~Harvey and G.~W.~Moore,
  {\it I-brane inflow and anomalous couplings on D-branes},
  Class.\ Quant.\ Grav.\  {\bf 14}, 47 (1997)
  [arXiv:hep-th/9605033].


\bibitem{Tseytlin:BI-DBI}
  A.~A.~Tseytlin,
  {\it Born-Infeld action, supersymmetry and string theory}, in {\it The many faces of the
  superworld. Yuri Golfand memorial volume}, M.A. Shifman, M.A.
  (ed.), World Scientific 2000,
  pp. 417-452
  [hep-th/9908105].




\bibitem{D-braneWZ}
  M.~B.~Green, C.~M.~Hull and P.~K.~Townsend,
  {\it D-Brane Wess-Zumino Actions, T-Duality and the Cosmological
  Constant},
  Phys.\ Lett.\  {\bf B382}, 65 (1996)
  [hep-th/9604119].



\bibitem{Tseytlin:DBInA}
  A.~A.~Tseytlin,
{\it On non-abelian generalisation of the Born-Infeld action in
string theory},
  Nucl.\ Phys.\ {\bf B501}, 41-52 (1997)
  [hep-th/9701125].




\bibitem{Dima01}
 D.~P.~Sorokin,
  {\it  Coincident (super)-Dp-branes of codimension one},
  JHEP {\bf 08}, 022 (2001)
  [hep-th/0106212];
  {\it Space-time symmetries and supersymmetry of coincident
  D-branes},
  Fortsch.\ Phys.\  {\bf 50}, 973 (2002);
 S.~Panda and D.~Sorokin,
  {\it Supersymmetric and kappa-invariant coincident D0-branes},
  JHEP {\bf 0302} (2003) 055
  [hep-th/0301065].
\bibitem{Drummond:2002kg}
  J.~M.~Drummond, P.~S.~Howe and U.~Lindstrom,
  {\it Kappa-symmetric non-Abelian Born-Infeld actions in three dimensions},
  Class.\ Quant.\ Grav.\  {\bf 19}, 6477 (2002)
  [hep-th/0206148].

\bibitem{IB+PKT=08-2}
  I.~A.~Bandos and P.~K.~Townsend,
''SDiff Gauge Theory and the M2 Condensate,''
 JHEP {\bf 0902}, 013 (2009)
  [arXiv:0808.1583 [hep-th]].


\bibitem{Marcus:1986cm}
  N.~Marcus and A.~Sagnotti,
  ``Group Theory From Quarks At The Ends Of Strings,''
  Phys.\ Lett.\  B {\bf 188} (1987) 58.

\bibitem{CremmerFerrara80}
E.~Cremmer and S.~Ferrara, {\it Formulation of eleven-dimensional
supergravity in superspace}, Phys. Lett. {\bf B91}, 61 (1980).


\bibitem{BrinkHowe80}
L.~Brink and P.S.~Howe, {\it Eleven-dimensional supergravity on the
mass-shell in superspace}, Phys. Lett. {\bf B91}, 384 (1980).


\bibitem{Howe+West=1983}
 P.~S.~Howe and P.~C.~West,
 ``The Complete N=2, D = 10 Supergravity",
 Nucl.\ Phys.\  {\bf B238}, 181-220 (1984).


\bibitem{Bellucci+Gates+89}
S.~Bellucci, S.~J.~J.~Gates, B.~Radak and S.~Vashakidze,
    {\it Improved supergeometries for type-II Green-Schwarz non-linear sigma-model},
  Mod.\ Phys.\ Lett.\  A {\bf 4}, 1985-1998 (1989).



\bibitem{stv}
  D.~P.~Sorokin, V.~I.~Tkach and D.~V.~Volkov,
  {\it Superparticles, twistors and Siegel symmetry},
  Mod.\ Phys.\ Lett.\  {\bf A4} (1989) 901-908.

\bibitem{DGHS93}
 F.~Delduc, A.~Galperin, P.~S.~Howe and E.~Sokatchev,
  {\it A Twistor formulation of the heterotic D$=$10 superstring with manifest
  (8,0) world sheet supersymmetry},
  Phys.\ Rev.\  {\bf D47}, 578 (1993)
  [hep-th/9207050];  see  \cite{Dima99} for more references on STV approach to superparticles and
  D<10 superstrings.


\bibitem{vz} D.~V.~Volkov and A.~A.~Zheltukhin,
  {\it Extension of the Pernose representation and its use to
  describe supersymmetric models},
  JETP Lett.\  {\bf 48} (1988) 63;
 {\it On the equivalence of the Lagrangians of massless Dirac and
 supersymmetrical particles},
  Lett.\ Math.\ Phys.\  {\bf 17}, 141 (1989);
 {\it Lagrangians For Massless Particles And Strings With Local And Global
  Supersymmetry},
  Nucl.\ Phys.\   {\bf B335}, 723 (1990).


\bibitem{stvz} D.~P.~Sorokin, V.~I.~Tkach, D.~V.~Volkov and A.~A.~Zheltukhin,
  {\it From the superparticle Siegel symmetry to the spinning
  particle proper time supersymmetry},
  Phys.\ Lett.\  {\bf B216}, 302 (1989).


\bibitem{Uvarov}
D.~V.~Uvarov,
  {\it On covariant kappa-symmetry fixing and the relation between the NSR  string
  and the type II GS superstring},
  Phys.\ Lett.\  {\bf B493}, 421-429 (2000)
  [arXiv:hep-th/0006185];
{\it New superembeddings for type II superstrings},
  JHEP {\bf 0207}, 008 (2002)
  [arXiv:hep-th/0112155].




\bibitem{Sok} E.~Sokatchev, {\it Light cone harmonic superspace and its
applications},
  Phys.\ Lett.\  {\bf B169}, 209-214 (1986);
  {\it Harmonic superparticle},
  Class.\ Quant.\ Grav.\  {\bf 4}, 237-246 (1987).


\bibitem{B90} I.~A.~Bandos,
  {\it A superparticle in Lorentz-harmonic superspace},
  Sov.\ J.\ Nucl.\ Phys.\  {\bf 51}, 906-914 (1990);

\bibitem{Ghsds}
  A.~S.~Galperin, P.~S.~Howe and K.~S.~Stelle, {\it The superparticle and the Lorentz
  group},  Nucl.\ Phys.\  {\bf B368}, 248-280 (1992)
  [hep-th/9201020];
\\  F.~Delduc, A.~Galperin and E.~Sokatchev,
  {\it Lorentz harmonic (super)fields and (super)particles},
  Nucl.\ Phys.\  B {\bf 368}, 143-171 (1992).


\bibitem{BZ-p} I.~A.~Bandos and A.~A.~Zheltukhin, ``Spinor Cartan moving N hedron, Lorentz harmonic formulations of
  superstrings, and kappa symmetry,''
  JETP Lett.\  {\bf 54}, 421-424 (1991);
``Green-Schwarz superstrings in spinor moving frame formalism'',
Phys.\ Lett.\ {\bf B288}, 77--84 (1992);
``Generalization of Newman-Penrose dyads in
connection with the action integral for supermembranes in an eleven-dimensional
space'',   JETP Lett.\ {\bf 55}, 81 (1992);
``D = 10 superstring:
Lagrangian and Hamiltonian mechanics in twistor-like Lorentz
harmonic formulation'', Phys.\ Part.\ Nucl.\ {\bf 25},
453-477 (1994);
 ``Eleven-dimensional supermembrane in a spinor
moving repere formalism'', Int.\ J.\ Mod.\ Phys.\ A {\bf 8}, 1081 (1993);
``N=1 super-p-branes in twistor - like Lorentz
harmonic formulation'', Class.\ Quant.\ Grav.\ {\bf 12}, 609 (1995)
[hep-th/9405113].



\bibitem{IB+DS06}
 I.~Bandos and D.~Sorokin,
  {\it Aspects of D-brane dynamics in supergravity backgrounds with fluxes,
  kappa-symmetry and equations of motion. IIB story},
  Nucl.\ Phys.\  {\bf B759}, 399 (2006)
  [hep-th/0607163].

\bibitem{Bandos:2008ba}
  I.~A.~Bandos,
  ``Superembedding approach to superstring in AdS(5)xS*5 superspace,''
  arXiv:0812.0257 [hep-th].




\bibitem{HoweSezgin04}
 P.~S.~Howe and E.~Sezgin,
  {\it The supermembrane revisited},
  Class.\ Quant.\ Grav.\  {\bf 22}, 2167-2200, (2005)
  [arXiv:hep-th/0412245].



\bibitem{5-equiv}
 P.S. Howe, E. Sezgin and P.C. West,
{\sl Phys.Lett.} {\bf B399}, 49--59 (1997), {\bf hep--th/9702008}; {\sl Phys.Lett.}
{\bf B400}, 255--259 (1997), {\bf hep--th/9702111};
\\ I. Bandos, K. Lechner, A. Nurmagambetov, P. Pasti, D. Sorokin and M.
Tonin, {\sl Phys.Lett.} {\bf B408}, 135-141 (1997), {\bf hep-th/9703127}.

\bibitem{hsw}
  P.~S.~Howe, E.~Sezgin and P.~C.~West,
  ``Covariant field equations of the M-theory five-brane,''
  Phys.\ Lett.\  {\bf B399}, 49 (1997)
  [arXiv:hep-th/9702008];
  ``The six-dimensional self-dual tensor,''
  Phys.\ Lett.\  {\bf B400}, 255 (1997)
  [arXiv:hep-th/9702111].


\bibitem{blnpst2}
  I.~A.~Bandos, K.~Lechner, A.~Nurmagambetov, P.~Pasti, D.~P.~Sorokin and M.~Tonin,
  ``On the equivalence of different formulations of the M theory  five-brane,''
  Phys.\ Lett.\  {\bf B408}, 135 (1997)
  [arXiv:hep-th/9703127].



\bibitem{Cederwall:2001td}
  M.~Cederwall, B.~E.~W.~Nilsson and D.~Tsimpis,
  ``D = 10 super-Yang-Mills at O(alpha**2),''
  JHEP {\bf 0107}, 042 (2001)
  [arXiv:hep-th/0104236].


\bibitem{Movshev:2009ba}
  M.~Movshev and A.~Schwarz,
  ``Supersymmetric Deformations of Maximally Supersymmetric Gauge Theories.
  I,''
  arXiv:0910.0620 [hep-th].




\bibitem{Cederwall:2001dx}
  M.~Cederwall, B.~E.~W.~Nilsson and D.~Tsimpis,
  ``Spinorial cohomology and maximally supersymmetric theories,''
  JHEP {\bf 0202}, 009 (2002)
  [arXiv:hep-th/0110069];  \\
  M.~Cederwall, B.~E.~W.~Nilsson and D.~Tsimpis,
  ``Spinorial cohomology of abelian d = 10 super-Yang-Mills at  O(alpha'**3),''
  JHEP {\bf 0211}, 023 (2002)
  [arXiv:hep-th/0205165];
  M.~Cederwall, U.~Gran, B.~E.~W.~Nilsson and D.~Tsimpis,
  ``Supersymmetric corrections to eleven-dimensional supergravity,''
  JHEP {\bf 0505}, 052 (2005)
  [arXiv:hep-th/0409107].

\bibitem{Bergshoeff:2001dc}
  E.~A.~Bergshoeff, A.~Bilal, M.~de Roo and A.~Sevrin,
  ``Supersymmetric non-abelian Born-Infeld revisited,''
  JHEP {\bf 0107}, 029 (2001)
  [arXiv:hep-th/0105274].



\bibitem{PKT+6d}
  T.~Kugo and P.~K.~Townsend,
  ``Supersymmetry And The Division Algebras,''
  Nucl.\ Phys.\  B {\bf 221}, 357--380 (1983); \\
  P.~S.~Howe, G.~Sierra and P.~K.~Townsend,
  ``Supersymmetry In Six-Dimensions,''
  Nucl.\ Phys.\  B {\bf 221}, 331-348 (1983);

\bibitem{Witten:1996hc}
  E.~Witten,
  ``Five-brane effective action in M-theory,''
  J.\ Geom.\ Phys.\  {\bf 22}, 103 (1997)
  [arXiv:hep-th/9610234] and refs. therein.



\bibitem{Dima+Eric=2007}
  E.~Bergshoeff, J.~Hartong and D.~Sorokin,
  {\it Q7-branes and their coupling to IIB supergravity},
  JHEP {\bf 0712}, 079 (2007)
[arXiv:0708.2287 [hep-th]].


\bibitem{Bagger:2007jr}
  J.~Bagger and N.~Lambert,
  {\it Gauge Symmetry and Supersymmetry of Multiple M2-Branes},
  Phys.\ Rev.\  D {\bf 77} (2008) 065008
  [arXiv:0711.0955 [hep-th]];
  {\it Comments On Multiple M2-branes},
  JHEP {\bf 0802} (2008) 105
  [arXiv:0712.3738 [hep-th]]; \\
  A.~Gustavsson,
  {\it Algebraic structures on parallel M2-branes},
  arXiv:0709.1260 [hep-th];
  {\it Selfdual strings and loop space Nahm equations},
  JHEP {\bf 0804}, 083 (2008)
  [arXiv:0802.3456 [hep-th]].

\bibitem{Aharony:2008ug}
  O.~Aharony, O.~Bergman, D.~L.~Jafferis and J.~Maldacena,
  {\it N=6 superconformal Chern-Simons-matter theories, M2-branes and their
  gravity duals},
  arXiv:0806.1218 [hep-th]; \\
  O.~Aharony, O.~Bergman and D.~L.~Jafferis,
  {\it Fractional M2-branes},
  JHEP {\bf 0811}, 043 (2008)
  [arXiv:0807.4924 [hep-th]].


\end{thebibliography}
}
 \end{document}